\pdfoutput=1
\documentclass[12pt,a4paper]{article}

\usepackage{multirow}

\usepackage{ifthen} \newboolean{pdflatex}
\setboolean{pdflatex}{true} 

\newboolean{articletitles}
\setboolean{articletitles}{true} 

\newboolean{uprightparticles}
\setboolean{uprightparticles}{false}

\def\paperauthors{LHCb collaboration} \def\paperasciititle{Precise measurement of the fs/fd ratio of fragmentation fractions and of Bs decay branching fractions} \def\papertitle{Precise measurement of the \fsfd ratio of fragmentation fractions\\ and of \Bs decay branching fractions} \def\paperkeywords{{High Energy Physics}, {LHCb}} \def\papercopyright{\the\year\ CERN for the benefit of the LHCb collaboration} \def\paperlicence{CC BY 4.0 licence}
\def\paperlicenceurl{https://creativecommons.org/licenses/by/4.0/}

\usepackage[top=1in, bottom=1.25in, left=1in, right=1in]{geometry}

\usepackage[section]{placeins}

\usepackage{microtype}
\usepackage{lineno}  \usepackage{xspace} \usepackage{caption}

\usepackage{graphicx}  \usepackage{color}
\usepackage{colortbl}
\graphicspath{{./figs/}} 

\usepackage{amsmath} \usepackage{amssymb}
\usepackage{amsfonts}
\usepackage{upgreek} 

\newcommand*\patchAmsMathEnvironmentForLineno[1]{\expandafter\let\csname old#1\expandafter\endcsname\csname #1\endcsname
\expandafter\let\csname oldend#1\expandafter\endcsname\csname
end#1\endcsname
 \renewenvironment{#1}{\linenomath\csname old#1\endcsname}{\csname oldend#1\endcsname\endlinenomath}}
\newcommand*\patchBothAmsMathEnvironmentsForLineno[1]{\patchAmsMathEnvironmentForLineno{#1}\patchAmsMathEnvironmentForLineno{#1*}}
\AtBeginDocument{\patchBothAmsMathEnvironmentsForLineno{equation}\patchBothAmsMathEnvironmentsForLineno{align}\patchBothAmsMathEnvironmentsForLineno{flalign}\patchBothAmsMathEnvironmentsForLineno{alignat}\patchBothAmsMathEnvironmentsForLineno{gather}\patchBothAmsMathEnvironmentsForLineno{multline}\patchBothAmsMathEnvironmentsForLineno{eqnarray}}

\usepackage{hyperxmp}

\usepackage[pdftex,
            pdfauthor={\paperauthors},
            pdftitle={\paperasciititle},
            pdfkeywords={\paperkeywords},
            pdfcopyright={Copyright (C) \papercopyright},
            pdflicenseurl={\paperlicenceurl}]{hyperref}

\usepackage[colorinlistoftodos,textsize=scriptsize]{todonotes}

\usepackage[bottom,flushmargin,hang,multiple]{footmisc}

\usepackage[all]{hypcap} 

\usepackage{xspace} 
\usepackage{upgreek}

\def\lhcb   {\mbox{LHCb}\xspace}

\def\bfactories {\mbox{\B Factories}\xspace}

\def\MagUp {\mbox{\em Mag\kern -0.05em Up}\xspace}

\ifthenelse{\boolean{uprightparticles}}{

 \def\Peta        {\ensuremath{\upeta}\xspace}

 \def\Pmu         {\ensuremath{\upmu}\xspace}                 
 \def\Pnu         {\ensuremath{\upnu}\xspace}                 
                  
 \def\Ppi         {\ensuremath{\uppi}\xspace}

 \def\Pphi        {\ensuremath{\upphi}\xspace}                 
                  
 \def\Pchi        {\ensuremath{\upchi}\xspace}                 
 \def\Ppsi        {\ensuremath{\uppsi}\xspace}

 \def\PDelta      {\ensuremath{\Delta}\xspace}                 
 \def\PXi         {\ensuremath{\Xi}\xspace}                 
 \def\PLambda     {\ensuremath{\Lambda}\xspace}                 
 \def\PSigma      {\ensuremath{\Sigma}\xspace}                 
 \def\POmega      {\ensuremath{\Omega}\xspace}                 
 \def\PUpsilon    {\ensuremath{\Upsilon}\xspace}

 \def\PB      {\ensuremath{\mathrm{B}}\xspace}                 
                  
 \def\PD      {\ensuremath{\mathrm{D}}\xspace}

 \def\PJ      {\ensuremath{\mathrm{J}}\xspace}                 
 \def\PK      {\ensuremath{\mathrm{K}}\xspace}

 \def\PX      {\ensuremath{\mathrm{X}}\xspace}

 \def\Pb      {\ensuremath{\mathrm{b}}\xspace}                 
 \def\Pc      {\ensuremath{\mathrm{c}}\xspace}                 
 \def\Pd      {\ensuremath{\mathrm{d}}\xspace}

 \def\Pi      {\ensuremath{\mathrm{i}}\xspace}

 \def\Pp      {\ensuremath{\mathrm{p}}\xspace}

 \def\Ps      {\ensuremath{\mathrm{s}}\xspace}                 
                  
 \def\Pu      {\ensuremath{\mathrm{u}}\xspace}

 \def\thebaroffset{0.0em}
}
{

 \def\Peta        {\ensuremath{\eta}\xspace}

 \def\Pmu         {\ensuremath{\mu}\xspace}                 
 \def\Pnu         {\ensuremath{\nu}\xspace}                 
                  
 \def\Ppi         {\ensuremath{\pi}\xspace}

 \def\Pphi        {\ensuremath{\phi}\xspace}                 
                  
 \def\Pchi        {\ensuremath{\chi}\xspace}                 
 \def\Ppsi        {\ensuremath{\psi}\xspace}                 
                  
 \mathchardef\PDelta="7101
 \mathchardef\PXi="7104
 \mathchardef\PLambda="7103
 \mathchardef\PSigma="7106
 \mathchardef\POmega="710A
 \mathchardef\PUpsilon="7107
                  
 \def\PB      {\ensuremath{B}\xspace}                 
                  
 \def\PD      {\ensuremath{D}\xspace}

 \def\PJ      {\ensuremath{J}\xspace}                 
 \def\PK      {\ensuremath{K}\xspace}

 \def\PX      {\ensuremath{X}\xspace}

 \def\Pb      {\ensuremath{b}\xspace}                 
 \def\Pc      {\ensuremath{c}\xspace}                 
 \def\Pd      {\ensuremath{d}\xspace}

 \def\Pi      {\ensuremath{i}\xspace}

 \def\Pp      {\ensuremath{p}\xspace}

 \def\Ps      {\ensuremath{s}\xspace}                 
                  
 \def\Pu      {\ensuremath{u}\xspace}

 \def\thebaroffset{0.18em}
}
\newcommand{\offsetoverline}[2][\thebaroffset]{\kern #1\overline{\kern -#1 #2}}

\makeatletter
\ifcase \@ptsize \relax \newcommand{\miniscule}{\@setfontsize\miniscule{4}{5}}\or \newcommand{\miniscule}{\@setfontsize\miniscule{5}{6}}\or \newcommand{\miniscule}{\@setfontsize\miniscule{5}{6}}\fi
\makeatother

\DeclareRobustCommand{\optbar}[1]{\shortstack{{\miniscule (\rule[.5ex]{1.25em}{.18mm})}
  \\ [-.7ex] $#1$}}

\def\mup        {{\ensuremath{\Pmu^+}}\xspace}

\def\mumu       {{\ensuremath{\Pmu^+\Pmu^-}}\xspace}

\def\neu        {{\ensuremath{\Pnu}}\xspace}

\def\neum       {{\ensuremath{\neu_\mu}}\xspace}

\def\uquark    {{\ensuremath{\Pu}}\xspace}

\def\dquark    {{\ensuremath{\Pd}}\xspace}

\def\squark    {{\ensuremath{\Ps}}\xspace}

\def\cquark    {{\ensuremath{\Pc}}\xspace}

\def\bquark    {{\ensuremath{\Pb}}\xspace}

\def\pion   {{\ensuremath{\Ppi}}\xspace}

\def\pip    {{\ensuremath{\pion^+}}\xspace}
\def\pim    {{\ensuremath{\pion^-}}\xspace}

\def\pimp   {{\ensuremath{\pion^\mp}}\xspace}

\def\kaon    {{\ensuremath{\PK}}\xspace}
\def\Kbar    {{\ensuremath{\offsetoverline{\PK}}}\xspace}

\def\KorKbar {\kern \thebaroffset\optbar{\kern -\thebaroffset \PK}{}\xspace}

\def\Kzb     {{\ensuremath{\Kbar{}^0}}\xspace}
\def\Kp      {{\ensuremath{\kaon^+}}\xspace}
\def\Km      {{\ensuremath{\kaon^-}}\xspace}
\def\Kpm     {{\ensuremath{\kaon^\pm}}\xspace}
\def\Kmp     {{\ensuremath{\kaon^\mp}}\xspace}
\def\KS      {{\ensuremath{\kaon^0_{\mathrm{S}}}}\xspace}

\def\Kstarz  {{\ensuremath{\kaon^{*0}}}\xspace}
\def\Kstarzb {{\ensuremath{\Kbar{}^{*0}}}\xspace}

\def\Kstarp  {{\ensuremath{\kaon^{*+}}}\xspace}
\def\Kstarm  {{\ensuremath{\kaon^{*-}}}\xspace}
\def\Kstarpm {{\ensuremath{\kaon^{*\pm}}}\xspace}

\newcommand{\etapr}{\ensuremath{\Peta^{\prime}}\xspace}

\def\Dbar    {{\ensuremath{\offsetoverline{\PD}}}\xspace}
\def\D       {{\ensuremath{\PD}}\xspace}

\def\DorDbar {\kern \thebaroffset\optbar{\kern -\thebaroffset \PD}\xspace}
\def\Dz      {{\ensuremath{\D^0}}\xspace}
\def\Dzb     {{\ensuremath{\Dbar{}^0}}\xspace}
\def\Dp      {{\ensuremath{\D^+}}\xspace}
\def\Dm      {{\ensuremath{\D^-}}\xspace}

\def\Dmp     {{\ensuremath{\D^\mp}}\xspace}
\def\DpDm    {\ensuremath{\Dp {\kern -0.16em \Dm}}\xspace}
\def\Dstar   {{\ensuremath{\D^*}}\xspace}

\def\Dstarzb {{\ensuremath{\Dbar{}^{*0}}}\xspace}

\def\Dstarm  {{\ensuremath{\D^{*-}}}\xspace}
\def\Dstarpm {{\ensuremath{\D^{*\pm}}}\xspace}
\def\Dstarmp {{\ensuremath{\D^{*\mp}}}\xspace}

\def\Ds      {{\ensuremath{\D^+_\squark}}\xspace}
\def\Dsp     {{\ensuremath{\D^+_\squark}}\xspace}
\def\Dsm     {{\ensuremath{\D^-_\squark}}\xspace}

\def\Dsmp    {{\ensuremath{\D^{\mp}_\squark}}\xspace}

\def\Dssp    {{\ensuremath{\D^{*+}_\squark}}\xspace}
\def\Dssm    {{\ensuremath{\D^{*-}_\squark}}\xspace}
\def\Dsspm   {{\ensuremath{\D^{*\pm}_\squark}}\xspace}

\def\B       {{\ensuremath{\PB}}\xspace}
\def\Bbar    {{\ensuremath{\offsetoverline{\PB}}}\xspace}

\def\BorBbar {\kern \thebaroffset\optbar{\kern -\thebaroffset \PB}\xspace}

\def\Bd      {{\ensuremath{\B^0}}\xspace}

\def\BdorBdbar {\kern \thebaroffset\optbar{\kern -\thebaroffset \Bd}\xspace}
\def\Bu      {{\ensuremath{\B^+}}\xspace}

\def\Bp      {{\ensuremath{\Bu}}\xspace}

\def\Bs      {{\ensuremath{\B^0_\squark}}\xspace}
\def\Bsb     {{\ensuremath{\Bbar{}^0_\squark}}\xspace}
\def\BsorBsbar {\kern \thebaroffset\optbar{\kern -\thebaroffset \Bs}\xspace}

\def\jpsi     {{\ensuremath{{\PJ\mskip -3mu/\mskip -2mu\Ppsi}}}\xspace}
\def\psitwos  {{\ensuremath{\Ppsi{(2S)}}}\xspace}

\def\chicone  {{\ensuremath{\Pchi_{\cquark 1}}}\xspace}

\def\Y#1S{\ensuremath{\PUpsilon{(#1S)}}\xspace}

\def\proton      {{\ensuremath{\Pp}}\xspace}
\def\antiproton  {{\ensuremath{\overline \proton}}\xspace}

\def\Lz          {{\ensuremath{\PLambda}}\xspace}
\def\Lbar        {{\ensuremath{\offsetoverline{\PLambda}}}\xspace}
\def\LorLbar     {\kern \thebaroffset\optbar{\kern -\thebaroffset \PLambda}\xspace}

\def\Lb           {{\ensuremath{\Lz^0_\bquark}}\xspace}

\def\BF         {{\ensuremath{\mathcal{B}}}\xspace}

\newcommand{\decay}[2]{\ensuremath{#1\!\to #2}\xspace} 

\def\to                 {\ensuremath{\rightarrow}\xspace}

\newcommand{\tauBs}{{\ensuremath{\tau_{\Bs}}}\xspace}
\newcommand{\tauBd}{{\ensuremath{\tau_{\Bd}}}\xspace}

\newcommand{\tauBu}{{\ensuremath{\tau_{\Bp}}}\xspace}

\def\qsq       {{\ensuremath{q^2}}\xspace}

\def\eps   {{\ensuremath{\varepsilon}}\xspace}

\def\Vud  {{\ensuremath{V_{\uquark\dquark}^{\phantom{\ast}}}}\xspace}

\def\Vus  {{\ensuremath{V_{\uquark\squark}^{\phantom{\ast}}}}\xspace}

\def\Vcb  {{\ensuremath{V_{\cquark\bquark}^{\phantom{\ast}}}}\xspace}

\def\BsToJPsiPhi  {\decay{\Bs}{\jpsi\phi}}

\def\AT#1     {\ensuremath{A_{\mathrm{T}}^{#1}}\xspace}

\def\C#1      {\ensuremath{\mathcal{C}_{#1}}\xspace}                       \def\Cp#1     {\ensuremath{\mathcal{C}_{#1}^{'}}\xspace}                    \def\Ceff#1   {\ensuremath{\mathcal{C}_{#1}^{\mathrm{(eff)}}}\xspace}        \def\Cpeff#1  {\ensuremath{\mathcal{C}_{#1}^{'\mathrm{(eff)}}}\xspace}       \def\Ope#1    {\ensuremath{\mathcal{O}_{#1}}\xspace}                       \def\Opep#1   {\ensuremath{\mathcal{O}_{#1}^{'}}\xspace}

\newcommand{\aunit}[1]{\ensuremath{\text{\,#1}}}

\newcommand{\tev}{\aunit{Te\kern -0.1em V}\xspace}
\newcommand{\gev}{\aunit{Ge\kern -0.1em V}\xspace}
\newcommand{\mev}{\aunit{Me\kern -0.1em V}\xspace}
\newcommand{\kev}{\aunit{ke\kern -0.1em V}\xspace}
\newcommand{\ev}{\aunit{e\kern -0.1em V}\xspace}
 
\newcommand{\mevc}{\ensuremath{\aunit{Me\kern -0.1em V\!/}c}\xspace}
\newcommand{\gevc}{\ensuremath{\aunit{Ge\kern -0.1em V\!/}c}\xspace}
\newcommand{\mevcc}{\ensuremath{\aunit{Me\kern -0.1em V\!/}c^2}\xspace}
\newcommand{\gevcc}{\ensuremath{\aunit{Ge\kern -0.1em V\!/}c^2}\xspace}

\def\pb {\aunit{pb}\xspace}
\def\invpb {\ensuremath{\pb^{-1}}\xspace}
\def\fb   {\ensuremath{\aunit{fb}}\xspace}
\def\invfb   {\ensuremath{\fb^{-1}}\xspace}

\def\ps   {\ensuremath{\aunit{ps}}\xspace}

\newcommand{\chisq}{\ensuremath{\chi^2}\xspace}

\def\gsim{{~\raise.15em\hbox{$>$}\kern-.85em
          \lower.35em\hbox{$\sim$}~}\xspace}
\def\lsim{{~\raise.15em\hbox{$<$}\kern-.85em
          \lower.35em\hbox{$\sim$}~}\xspace}

\def\sqs   {\ensuremath{\protect\sqrt{s}}\xspace}

\def\pt         {\ensuremath{p_{\mathrm{T}}}\xspace}

\def\evtgen     {\mbox{\textsc{EvtGen}}\xspace}

\def\geant      {\mbox{\textsc{Geant4}}\xspace}

\def\photos     {\mbox{\textsc{Photos}}\xspace}

\def\pythia     {\mbox{\textsc{Pythia}}\xspace}

\def\tell1  {TELL1\xspace}
\def\ukl1   {UKL1\xspace}

\newcommand{\eg}{\mbox{\itshape e.g.}\xspace}

\newcommand{\fsfd}{{\ensuremath{f_s/f_d}}\xspace}

\newcommand{\fsfu}{{\ensuremath{f_s/f_u}}\xspace}
\newcommand{\fsfufd}{{\ensuremath{f_s/(f_u+f_d)}}\xspace}
\newcommand{\fsfdorfu}{{\ensuremath{f_s/f_{d (u)}}}\xspace}

\def\fsfufdfrac{\ensuremath{\frac{f_s}{f_u+f_d}}\xspace}

\def\fsfufrac{\ensuremath{\frac{f_s}{f_u}}\xspace}
\def\fsfdfrac{\ensuremath{\frac{f_s}{f_d}}\xspace}
\def\fsfdorfufrac{\ensuremath{\frac{f_s}{f_{d(u)}}}\xspace}
\def\R{\ensuremath{\mathcal{R}}\xspace}

\newcommand{\xis}{{\ensuremath{\xi_{s}}\xspace}}

\def\BsDsorDsstmunu   {\decay{\Bs}{\D^{(*)-}_s \mup\neum}}
\def\BsDsXmunu   {\decay{\Bs}{\Dsm\PX\mup\neum}}
\def\BsDKXmunu   {\decay{\Bs}{\Dbar\Kbar\PX\mup\neum}}
\def\BsDKpXmunu   {\decay{\Bs}{\Dzb\Km\PX\mup\neum}}
\def\BsDKzXmunu   {\decay{\Bs}{\Dm\Kzb\PX\mup\neum}}

\def\BorDuXmunu    {\decay{\B^{+,0} }{\Dzb\PX\mup\neum}}
\def\BorDdXmunu    {\decay{\B^{+,0} }{\Dm\PX\mup\neum}}

\def\BDsKXmunu   {\decay{\B^{+,0} }{\Dsm\Kbar\PX\mup\neum}}

\newcommand{\ncorr}{{\ensuremath{n_{\mathrm{ corr}}}}\xspace}

\def\BsDspi   {\decay{\Bs}{\Dsm\pip}}
\def\BdDdK    {\decay{\Bd}{\Dm \Kp}}
\def\BdDpi    {\decay{\Bd}{\Dm \pip}}
\def\BdDdstK    {\decay{\Bd}{\Dstarm \Kp}}
\def\BdDstpi    {\decay{\Bd}{\Dstarm \pip}}
\def\BsDsK    {\decay{\Bs}{\Dsmp\Kpm}}
\def\BsDspipipi   {\decay{\Bs}{\Dsm\pip\pim\pip}}

\def\DsKKpi     {\decay{\Dsm}{\Km\Kp\pim}}
\def\DdKpipi    {\decay{\Dm}{\Kp\pim\pim}}
\def\DuKpi      {\decay{\Dzb}{\Kp\pim}}

\newcommand{\PhiPS}{\ensuremath{\Phi_{\mathrm{PS}}}\xspace}
\newcommand{\PhiPSDdK}{\ensuremath{\Phi_{\mathrm{PS},\Dm\Kp }}\xspace}
\newcommand{\PhiPSDpi}{\ensuremath{\Phi_{\mathrm{PS},\Dm\pip}}\xspace}

\newcommand{\Na}{\ensuremath{\mathcal{N}_a}\xspace}
\newcommand{\NF}{\ensuremath{\mathcal{N}_F}\xspace}
\newcommand{\NE}{\ensuremath{\mathcal{N}_E}\xspace}

\newcommand{\FR}{\ensuremath{\mathcal{F}_R}\xspace}

\def\BuToJPsiK {\decay{\Bu}{\jpsi\Kp}}
\def\bujpsik{\BuToJPsiK}
\def\bsdspi{\BsDspi}

\def\bsjpsiphi{\BsToJPsiPhi}

\def\BsToJPsiPhiKK  {\ensuremath{\bsjpsiphi, \Pphi\to \PK^+\PK^-}\xspace}
\def\bsjpsikk{\decay{\Bs}{\jpsi \Kp \Km}}
\def\bdjpsikstar{\decay{\Bd}{\jpsi \Kstarz}\xspace}
\def\bsmumu{\decay{\Bs}{\mumu}\xspace}
\def\chisquare{\ensuremath{\chi^2}\xspace}

\def\bsphiphi    {\decay{\Bs}{\phi\phi}}
\def\bsjpsipipi  {\decay{\Bs}{\jpsi\pip\pim}}

\def\Dsonem   {{\ensuremath{\D_{\squark 1}(2536)^-}}\xspace}

\definecolor{webgreen}{rgb}{0,.5,0}
\definecolor{webbrown}{rgb}{.6,0,0}
\definecolor{RoyalBlue}{rgb}{0,0,.5}

\def\porpbar {\kern \thebaroffset\optbar{\kern -\thebaroffset \Pp}\xspace}

\def\normcheck {{\ensuremath{\star}}\xspace} 
 \usepackage{booktabs}
\usepackage{cite} \usepackage{mciteplus}
 
\usepackage{longtable} \usepackage{changepage}

\begin{document}

\renewcommand{\thefootnote}{\fnsymbol{footnote}}
\setcounter{footnote}{1}

\begin{titlepage}
\pagenumbering{roman}

\vspace*{-1.5cm}
\centerline{\large EUROPEAN ORGANIZATION FOR NUCLEAR RESEARCH (CERN)}
\vspace*{1.5cm}
\noindent
\begin{tabular*}{\linewidth}{lc@{\extracolsep{\fill}}r@{\extracolsep{0pt}}}
\ifthenelse{\boolean{pdflatex}}{\vspace*{-1.5cm}\mbox{\!\!\!\includegraphics[width=.14\textwidth]{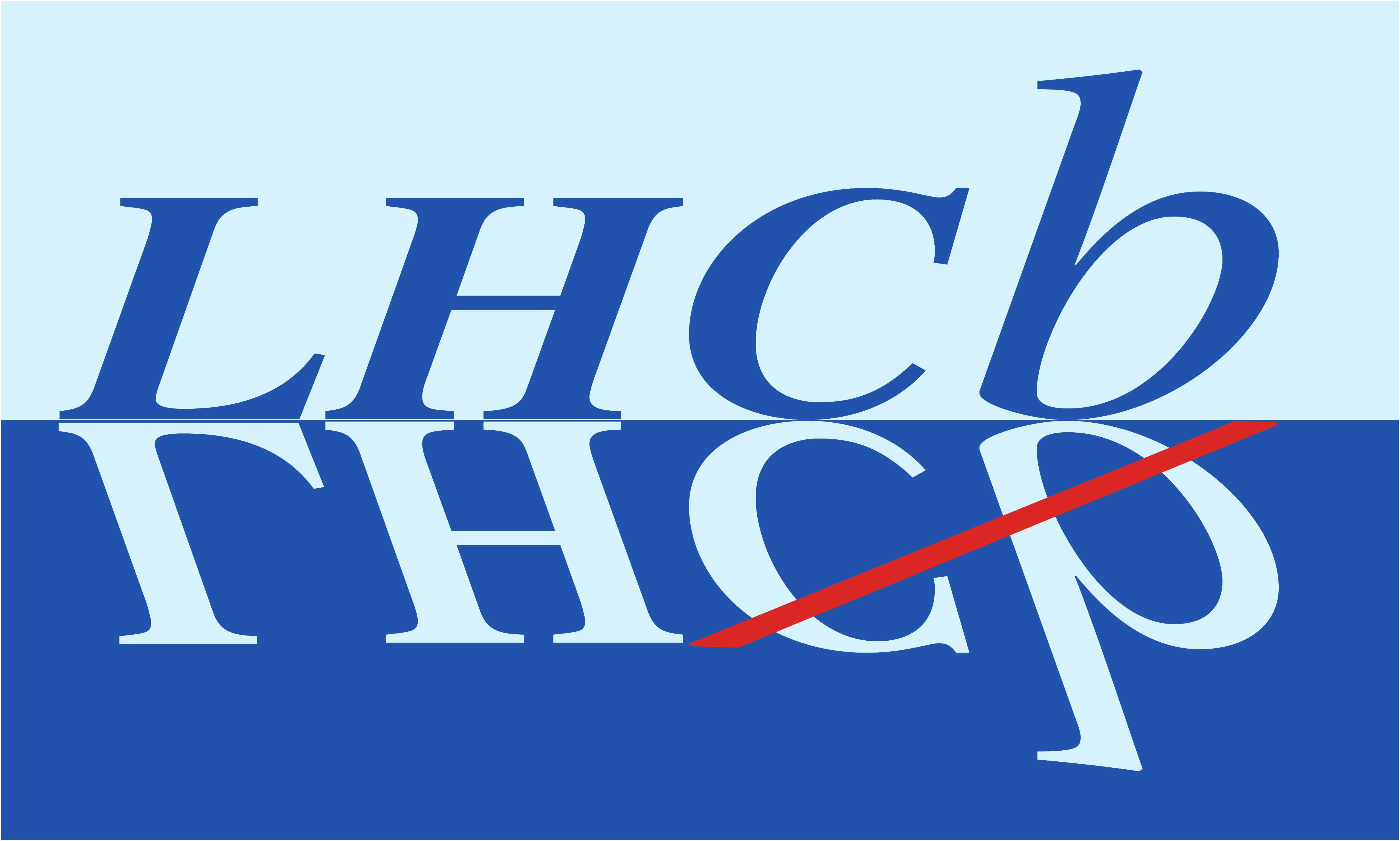}} & &}{\vspace*{-1.2cm}\mbox{\!\!\!\includegraphics[width=.12\textwidth]{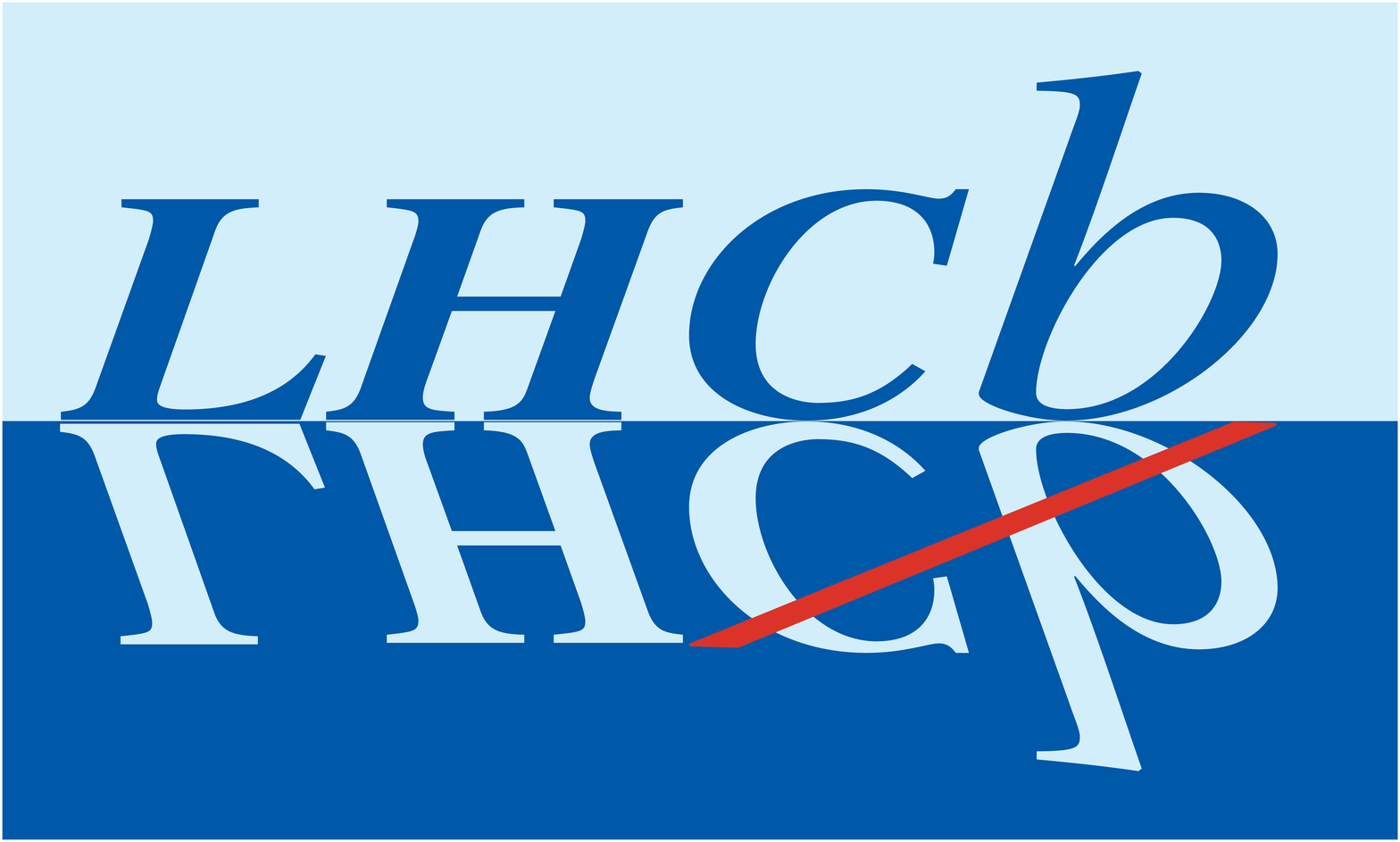}} & &}\\
 & & CERN-EP-2021-027 \\  & & LHCb-PAPER-2020-046 \\  & & September 2, 2021 \\ & & \\
\end{tabular*}

\vspace*{4.0cm}

{\normalfont\bfseries\boldmath\huge
\begin{center}
\papertitle 
\end{center}
}

\vspace*{2.0cm}

\begin{center}
\paperauthors\footnote{Authors are listed at the end of this paper.}
\end{center}

\vspace{\fill}

\begin{abstract}
  \noindent
The ratio of the \Bs and \Bd fragmentation fractions, \fsfd, in proton-proton collisions at the LHC, is obtained as a function of $\B$-meson transverse momentum 
and collision centre-of-mass energy from the combined analysis of different $\B$-decay channels 
measured by the LHCb experiment. 
The results are described by a linear function of the meson transverse momentum, 
or with a function inspired by Tsallis statistics. 
Precise measurements of the branching fractions of the \bsjpsiphi and \bsdspi decays are performed, reducing their uncertainty by about a factor of two 
with respect to previous world averages.
Numerous \Bs decay branching fractions, measured at the LHCb experiment, are also updated using the new values of \fsfd and branching fractions of normalisation channels. 
These results reduce a major source of systematic uncertainty in several searches for new physics performed through measurements of \Bs branching fractions. 
  \end{abstract}

\vspace{\fill}

\begin{center}
  Published in Phys.~Rev.~D104~(2021)~032005
\end{center}

\vspace{\fill}

{\footnotesize 
\centerline{\copyright~\papercopyright. \href{\paperlicenceurl}{\paperlicence}.}}
\vspace*{2mm}

\end{titlepage}

\newpage
\setcounter{page}{2}
\mbox{~}

\renewcommand{\thefootnote}{\arabic{footnote}}
\setcounter{footnote}{0}

\cleardoublepage

\pagestyle{plain} \setcounter{page}{1}
\pagenumbering{arabic}


\def\taueffbsjpsiphi{1.487}

\def\chsqvalue{133/61
 }
\def\fsfdseventev{(0.244 \pm 0.008)  + ((-10.3 \pm 2.7)\times 10^{-4})\cdot \pt
 }
\def\fsfdeighttev{(0.240 \pm 0.008)  + ((-\phantom{0}3.4 \pm 2.3)\times 10^{-4})\cdot \pt
 }
\def\fsfdthirteentev{(0.263 \pm 0.008)  + ((-17.6 \pm 2.1)\times 10^{-4})\cdot \pt
 }

\def\fsfdintseventev{\text{$0.2390 \pm 0.0076$
 }}
\def\fsfdinteighttev{\text{$0.2385 \pm 0.0075$
 }}
\def\fsfdintthirteentev{\text{$0.2539 \pm 0.0079$
 }}

\def\FRvalue{\ensuremath{0.505 \pm 0.016}
 }
\def\BRBsjpsiphikk{\text{$(5.01 \pm 0.16 \pm 0.17)\times 10^{-4}$
 }}
\def\BRBsjpsiphi{\text{$(1.018 \pm 0.032 \pm 0.037)\times 10^{-3}$
 }}

\def\RHBRwDvalue{\text{ $0.686 \pm 0.021$
  }}
\def\RHBRvalue{\text{ $1.18 \pm 0.04$
  }}
\def\BRBsDspivalue{\text{ $(3.20 \pm 0.10 \pm 0.16)\times 10^{-3}$
  }}

\section{Introduction}
\label{sec:Introduction}
Measurements of branching fractions of \Bs meson decays are sensitive tools to test the Standard Model~(SM) of particle physics.
They often require knowledge of the \Bs production rate. 
To avoid uncertainties related to the $b$-hadron production cross-section and integrated luminosity, 
and to partly cancel those related to detection efficiencies, 
at hadron colliders the \Bs branching fractions are often measured relative to other \B-meson decay channels. 
In the absence of any precisely known \Bs branching fraction, most measurements are normalised to \Bu or \Bd meson decays, 
and thus require the ratio of their fragmentation fractions as input.
The fragmentation fractions, denoted as $f_u$, $f_d$, $f_s$, and $f_{\rm baryon}$, are the probabilities for a \bquark quark to hadronise into 
a \Bu, \Bd, \Bs meson or a \bquark baryon.\footnote{The inclusion of the charge-conjugate modes is implied throughout this paper.}
These fractions include contributions from intermediate states decaying to the aforementioned hadrons via the strong or electromagnetic interaction. 

The \bquark-hadron fragmentation fractions in proton-proton (\proton\proton) collisions at the Large Hadron Collider (LHC) energies are in general different from 
those measured at $e^+e^-$ colliders~\cite{ACTON1992357,Buskulic:1995bd,Acciarri:435659,Abdallah:2003xp} 
or in $\proton\antiproton$ collisions at the Tevatron~\cite{Aaltonen:2008zd}, with which they were previously averaged~\cite{HFLAV18,PDG2020}. 
The ratios of fragmentation fractions are found to depend on kinematics, in particular on the \bquark-hadron transverse momentum with respect to the beam direction (\pt);
the dependence on the \bquark-hadron pseudorapidity ($\eta$) has also been studied, but not found to be significant~\cite{Aaltonen:2008zd,LHCb-PAPER-2011-018,LHCb-PAPER-2012-037}.
The ratio of fragmentation fractions \fsfu has also been shown to depend on the $pp$ collision centre-of-mass energy, \sqs~\cite{LHCb-PAPER-2019-020}.
In the following, $f_u = f_d$ is assumed to hold due to isospin symmetry.

The \bsjpsiphi decay is among the most studied of the \Bs-meson decays, owing to its relative abundance
and high reconstruction efficiency. As such, this decay is used as the normalisation channel for several other \Bs 
decays~\cite{LHCb-PAPER-2012-005,LHCb-PAPER-2012-010,LHCb-PAPER-2012-024,LHCb-PAPER-2015-023,LHCb-PAPER-2015-033}.
Despite this, the precision on its branching fraction is still limited; 
the most precise measurement was performed by the LHCb experiment
with $pp$ collision data collected at $\sqs=7\tev$, corresponding to an integrated luminosity of $1\invfb$. This measurement yields  \mbox{$\BF(\bsjpsiphi) = (1.050 \pm 0.013 \pm 0.064 \pm 0.082 )\times 10^{-3}$}~\cite{LHCb-PAPER-2012-040},
where the first uncertainty is statistical, the second systematic, 
including the external branching fraction measurement of \bujpsik decays,
and the third is due to the measurement of \fsfd~\cite{LHCb-PAPER-2011-018}. 
Other measurements were performed by the Belle~\cite{Thorne:2013llu} and CDF~\cite{Abe:1996kc} collaborations. 

The \BsDspi decay is another important \Bs meson decay mode,
which is used as the normalisation channel for several hadronic \Bs decays with a single charm meson in the final state;
its branching fraction can be used to test for the presence of physics beyond the SM in tree-level hadronic \B decays~\cite{Bordone:2020gao}.
However, the current precision on its branching fraction is also limited;
the current best measurement by the LHCb experiment was performed 
using $pp$ collision data collected at $\sqs=7\tev$, corresponding to $0.37\invfb$ of integrated luminosity.
This measurement yields \mbox{$\BF(\BsDspi) = (2.95 \pm 0.05 \pm 0.17 _{\,-\,0.22}^{\,+\,0.18})\times 10^{-3}$}~\cite{LHCb-PAPER-2011-022}, where 
the first uncertainty is statistical, the second systematic, 
including the external branching fraction measurement of \BdDpi decays,
and the third due to the measurement of \fsfd taken from Ref.~\cite{LHCb-PAPER-2011-018}.
Other measurements were performed by the Belle~\cite{Louvot:2008sc} and CDF~\cite{Abulencia:2006aa} collaborations.

The knowledge of \Bs branching fractions is thus often limited by the precision of the fragmentation fraction ratios.
This paper presents a simultaneous determination of the fragmentation fractions
and \Bs branching fractions with different decay modes. 
A combined analysis of LHCb measurements sensitive to \fsfd is performed,
in order to determine a precise value of this ratio as a function of \sqs and \pt as well as the \bsjpsiphi and \bsdspi branching fractions. 
This analysis employs previous LHCb measurements performed with ratios of semileptonic decays 
 $B\to \Dbar X \mup \neum $ at $\sqs=7\tev$~\cite{LHCb-PAPER-2011-018} and $13\tev$~\cite{LHCb-PAPER-2018-050}, where $X$ denotes possible additional particles,
 hadronic  $B\to D h$ decays, where $h = \pi, K$, at $\sqs=7$, 8 and $13\tev$~\cite{LHCb-PAPER-2012-037, LHCb-PAPER-2020-021}, 
 and  $B\to J/\psi h^{'}$ decays, where $h^{'} = K, \phi$,  at $\sqs=7$, 8 and $13\tev$~\cite{LHCb-PAPER-2019-020}.
 Measurements at 7 and 8 TeV were performed with data taken in 2010, 2011 and 2012, during Run 1 of the LHC; measurements at $13 \tev$ were performed with data taken in 2015 and 2016, during Run 2 of the LHC.
Combinations of the Run~1 measurements were performed in Refs.~\cite{LHCb-PAPER-2012-037,LHCb-CONF-2013-011} and are superseded by this paper.

This paper is organised as follows: in Sect.~\ref{sec:Measurements} the LHCb detector and the measurements used in this analysis are presented, along with their sensitivities to the fragmentation fractions and branching fractions. 
The combined fit to the data is introduced in Sect.~\ref{sec:Combined_fit}. The results of the fit for the differential and integrated fragmentation fractions and for the \bsjpsiphi and \bsdspi branching fractions are presented in Sect.~\ref{sec:Results}. 
In Sect.~\ref{sec:Updated_BFs}, these results are used to update about sixty different \Bs branching fractions measured so far by the LHCb experiment.
In Sect.~\ref{sec:Tsallis}, the data is also described by a function inspired by the Tsallis statistics. 
Finally, conclusions are drawn in Sect.~\ref{sec:Conclusion}.

\section{Measurements}
\label{sec:Measurements}

The LHCb detector~\cite{LHCb-DP-2008-001,LHCb-DP-2014-002} is a
single-arm forward spectrometer covering the pseudorapidity range $2 < \eta < 5$, designed for
the study of particles containing \bquark\ or \cquark\ quarks.
Simulation is used to model the effects of the detector acceptance and the imposed selection requirements.
 In the simulation, $pp$ collisions are generated using \pythia~\cite{Sjostrand:2007gs} 
 with a specific \lhcb configuration~\cite{LHCb-PROC-2010-056}.  Decays of unstable particles
  are described by \evtgen~\cite{Lange:2001uf}, in which final-state radiation is generated using \photos~\cite{davidson2015photos}.
  The interaction of the generated particles with the detector, and its response, are implemented using the \geant
  toolkit~\cite{Allison:2006ve, *Agostinelli:2002hh} as described in Ref.~\cite{LHCb-PROC-2011-006}. 

The five sets of measurements by the LHCb experiment~\cite{LHCb-PAPER-2011-018,LHCb-PAPER-2012-037,LHCb-PAPER-2018-050,LHCb-PAPER-2019-020,LHCb-PAPER-2020-021} that are combined in this paper rely on three different final states,
referred to as semileptonic, hadronic, and charmonium final states.
They are used to determine the ratio of efficiency-corrected yields, \ncorr, of $\Bs\to Y$ decays relative to \Bu or $\Bd \to Z$ decays,
which is sensitive to the ratio of branching fractions, \BF, multiplied by \fsfdorfu, 
\begin{equation}
	\frac{\ncorr(\Bs\to Y)}{\ncorr(\B^{0 (+)} \to Z)} = \frac{\BF(\Bs \to Y)}{\BF(\B^{0(+)} \to Z)} \fsfdorfufrac \quad ,
\label{eq:fsfd_general}
\end{equation}
where $\mathcal{B}$ is the exclusive branching fraction  for the hadronic and charmonium measurements, and the inclusive one for the semileptonic measurement.
The five sets of measurements and their sensitivity to fragmentation fractions and branching fractions are summarized in Table~\ref{tab:measurement_summary}.

\begin{table}[tp]
    \begin{center}
    \caption{Five sets of measurements by the LHCb experiment combined in this paper and their sensitivity to fragmentation fractions and branching fractions.}
    \label{tab:measurement_summary}

    \scalebox{0.85}{
    \begin{tabular}{l | c c c c  c }
    \toprule 
	    Final state & \sqs  & Relative or absolute & Sensitivity & Reference  \\
	        \midrule
	    $B\to \Dbar X \mup \neum $ & $7\tev$  & Absolute & \fsfd & \cite{LHCb-PAPER-2011-018}\\
	 \midrule    
	    $B\to \Dbar X \mup \neum $   & $13\tev$ & Absolute & \fsfd & \cite{LHCb-PAPER-2018-050}\\
    \midrule
	    $\BsDspi,\BdDdK$            & $7 \tev$ & Absolute & \fsfd & \cite{LHCb-PAPER-2012-037}    \\
	    $\BsDspi,\BdDpi$           & $7 \tev$ & Relative & \fsfd & \cite{LHCb-PAPER-2012-037}    \\
	 \midrule
	    $\BsDspi,\BdDpi$             & $7, 8, 13\tev$ & Absolute & \fsfd, $\BF(\BsDspi)$ & \cite{LHCb-PAPER-2020-021}    \\
    \midrule
	   $\bsjpsiphi,\bujpsik$          & $7, 8, 13\tev$ & Relative & $\fsfd$, $\BF(\bsjpsiphi)$ & \cite{LHCb-PAPER-2019-020} \\
        \bottomrule
    \end{tabular} }
    \end{center}
\end{table}

The various measurements have different ranges in pseudorapidity and transverse momentum of the \B meson. 
The semileptonic and hadronic measurements are performed for $\eta \in [ 2, 5]$, 
while the charmonium measurement extends this range to $\eta \in [ 2, 6.4 ]$.
As no pseudorapidity dependence is seen in the measurements under consideration, 
the fiducial region in which the combined analysis 
is considered valid includes the latter range. 
The combined analysis is performed as a function of \pt in the widest of the individual ranges, $\pt \in [0.5, 40] \gevc$, which is used in the charmonium 
measurement; it is maintained as the fiducial region.
The semileptonic measurement is performed for $\pt \in [4,25] \gevc$
and the hadronic measurement for $\pt \in [1.5,40] \gevc$.

The semileptonic measurements~\cite{LHCb-PAPER-2011-018,LHCb-PAPER-2018-050} use inclusive $B\to \Dbar X \mup \neum$ decays, 
having reconstructed a ground state charm meson and a muon.
The decay width of $\bquark \to \uquark$ decays is expected to be approximately 1\%~\cite{PDG2020} of the total semileptonic width 
and almost equal for \Bs, \Bd and \Bu mesons and is thus ignored.
The modes studied are \BsDsXmunu, \BsDKXmunu for the \Bs meson and \BorDuXmunu and \BorDdXmunu for the \Bu and \Bd mesons, the contributions of which are not separated.
As the \BsDKzXmunu final state cannot be reconstructed with high efficiency at the LHCb experiment, its contribution is inferred from the \BsDKpXmunu rate and the known decay modes of excited \Ds mesons to \D\kaon and \Dstar\kaon final states. 
The charm mesons are reconstructed using the decays \DsKKpi, \DdKpipi and \DuKpi.
The inclusive semileptonic decay widths for \Bs, \Bd and \Bu mesons are almost equal, 
apart from an SU(3) breaking correction factor of $1-\xis = 1.010 \pm 0.005$~\cite{Bigi_2011}, and are normalised to the 
corresponding total widths through the ratio of \Bs over \Bu and \Bd lifetimes, 
denoted as \tauBs, \tauBu and \tauBd.
Accordingly, \fsfufd is determined as
\begin{equation}
\begin{aligned}
\fsfufdfrac = &\frac{\ncorr(\BsDsXmunu)+\ncorr(\BsDKXmunu)}
	       {\ncorr(\BorDuXmunu) +\ncorr(\BorDdXmunu)} \frac{\tauBu+\tauBd}{2\tauBs}(1-\xis) \quad \\
	       -& \eps_{\mathrm{ratio}} \frac{\BF(\BDsKXmunu)}{ \BF_{\mathrm{SL} }   }   \quad ,
\label{eq:fsfd_sl}
\end{aligned}
\end{equation}
where the efficiency-corrected yields, \ncorr, incorporate the relevant charm-meson branching fractions.
The second term is small and is included to subtract the components from \BDsKXmunu decays which are reconstructed in the \BsDsXmunu sample, and contains $\eps_\mathrm{ratio}$, which is the ratio of efficiencies of reconstructing \BsDsXmunu and \BDsKXmunu through reconstruction of the $\Dsm\mup$ pair, and $\BF_{\mathrm{SL}}$, which is the semileptonic branching fraction of \Bs mesons\cite{LHCb-PAPER-2018-050}.
The efficiency-corrected yields have been corrected for cross-feeds;
\eg those in the denominator have had cross-feed contributions, from $\Bs,\Lb \to \Dbar X \mup \neum$ decays, subtracted.
The Run~1 measurement determines the integrated\footnote{Throughout this text, integrated \fsfd or \fsfufd refer to measurements integrated over \B-meson kinematics.} value of \fsfufd at $\sqs = 7 \tev$ using a data sample corresponding to an integrated luminosity of $3\invpb$~\cite{LHCb-PAPER-2011-018}.
The Run~2 measurement determines the value of \fsfufd in intervals of \B-meson \pt at $\sqs = 13 \tev$ using data corresponding to an integrated luminosity of $1.7\invfb$~\cite{LHCb-PAPER-2018-050}.

The hadronic measurements~\cite{LHCb-PAPER-2012-037, LHCb-PAPER-2020-021} make use of \BdDpi, \BdDdK and \BsDspi decays,
using the same decay modes for the charm mesons as for the semileptonic analysis (\DsKKpi and \DdKpipi).
As the ratio of branching fractions of the \BsDspi decay relative to $\Bd\to D^-h^+$ decays is predicted~\cite{Fleischer:2010ay,Fleischer:2010ca},
\fsfd can be determined according to
\begin{subequations}
\begin{align}
\fsfdfrac &= \PhiPSDdK \left| \frac{\Vus}{\Vud} \right|^2 \left( \frac{f_K}{f_\pi} \right)^2
		 \frac{\tauBd}{\tauBs} \frac{1}{\Na\NF} \frac{\BF(\DdKpipi)}{\BF(\DsKKpi)}
		 \frac{\ncorr(\BsDspi)}{\ncorr(\BdDdK)}  ~,  
\label{eq:fsfd_hadr} \\
	\fsfdfrac &= \PhiPSDpi  
		 \frac{\tauBd}{\tauBs} \frac{1}{\Na\NF\NE} \frac{\BF(\DdKpipi)}{\BF(\DsKKpi)}
		 \frac{\ncorr(\BsDspi)}{\ncorr(\BdDpi)}  \quad,
\label{eq:fsfd_hadr_dpi}
\end{align}
\end{subequations}
where \PhiPS is a phase-space factor, \Vus and \Vud are the Cabibbo–Kobayashi–Maskawa (CKM) matrix elements, and $f_K$ and $f_\pi$ are the kaon and pion decay constants, which have permille uncertainties~\cite{PDG2020}.
The remaining factors describe corrections to this ratio from non-factorisable effects, \Na, the form factors, \NF, and exchange diagram contributions to the \BdDpi decay, \NE.
The hadronic Run 1 measurement in Ref.~\cite{LHCb-PAPER-2012-037} uses a data sample corresponding to an integrated luminosity of $1\invfb$ at $\sqs = 7 \tev$ and determines both ratios in Eq.~\eqref{eq:fsfd_hadr} and~\eqref{eq:fsfd_hadr_dpi}.
The integrated value of \fsfd is determined using Eq.~\eqref{eq:fsfd_hadr};
the \pt dependence of \fsfd is determined in intervals of \pt using Eq.~\eqref{eq:fsfd_hadr_dpi}.
These results are included in a single dataset by scaling the \pt dependent measurement with the \Dm\pip final state to the integrated value of \fsfd measured with the \Dm\Kp final state.
The hadronic ratio measurement in Ref.~\cite{LHCb-PAPER-2020-021} 
uses data samples corresponding to integrated luminosities of $1\invfb$, $2\invfb$ and $2\invfb$ at $\sqs = 7$, $8$ and $13 \tev$, respectively,
to determine the ratio with \Dm\pip final state in Eq.~\eqref{eq:fsfd_hadr_dpi}, which is sensitive to the integrated value for \fsfd at each collision energy.

The charmonium measurement determines the \pt dependence of \fsfu at $\sqs = 7$, $8$ and $13 \tev$ using data samples corresponding 
to integrated luminosities of $1\invfb$, $2\invfb$ and $1.4\invfb$, respectively~\cite{LHCb-PAPER-2019-020}. 
It uses the decay modes \bsjpsiphi and \bujpsik, where the \Pphi meson decays to $\Kp\Km$, and leads to
\begin{equation}
\label{eq:R}
	\fsfufrac = \frac{\ncorr(\bsjpsiphi)}{\ncorr(\bujpsik)}  \frac{\BF (\bujpsik)} {\BF (\bsjpsiphi)\BF(\phi \to K^+ K^-) } = \frac{\R}{\FR} \quad, 
\end{equation}
where \R is the ratio of efficiency-corrected yields and \FR denotes the ratio of branching fractions.
As no prediction is available for the ratio \FR,  
this is included as a free parameter in the fit and is an additional result from this analysis.\footnote{In a measurement by the ATLAS collaboration~\cite{Aad:2015cda} 
the ratio \R was converted to a value for \fsfd using a prediction for the ratio of the \bsjpsiphi and \bdjpsikstar branching fractions~\cite{Liu:2013nea}.
In this paper, results from Ref.~\cite{Liu:2013nea} are not used because of  disputed theoretical uncertainties arising 
from the assumption of factorisation.}
The ratio \FR is therefore constrained in this measurement by the overall scale of \fsfd through the information provided by the analysis of the other final states; however, the large yield of this decay mode is very powerful for studying the \sqs and \pt dependence of the fragmentation fraction ratio.
The measurement in Ref.~\cite{LHCb-PAPER-2012-040} includes a full amplitude analysis of
the \bsjpsikk decay in order to separate the components in the $\PK^+\PK^-$ spectrum. 
The largest resonant contributions are from the $f_0(980)$, the $\phi$, and the $f^\prime(1525)$ mesons. 
In the mass region close to the $\Pphi$ resonance, in addition to the $f_0(980)$ meson, there is also a non-resonant S-wave component.
The total S-wave fraction is in general not negligible~\cite{LHCb-PAPER-2012-040} and varies as a function of the $\Kp\Km$ invariant mass.
When considering a small window around the \Pphi resonance mass, the S-wave contribution is significantly reduced. 
The \bsjpsiphi measurement from Ref.~\cite{LHCb-PAPER-2019-020}, required a tight mass window of $\pm 10 \mev$ around the \Pphi mass;
therefore, the contribution of the S-wave component is suppressed to $(1.0\pm 0.2)\%$. This contribution is subtracted from the final value of the branching fraction reported in this paper.

To determine \fsfd, the semileptonic and hadronic measurements rely on external inputs from theory and experiment;
most prominently, the \Dm, \Dzb and \Dsm meson branching fractions to the considered decay modes, the \Bu, \Bd and \Bs meson lifetimes,
and the theory predictions for the \Na, \NF, and \NE parameters.
In this combined analysis, all of the external inputs have been updated to their currently best known values, as shown in Table~\ref{tab:external_inputs}.
For $\BF(\DsKKpi)$, a recent result from BESIII~\cite{BESIII} is included and the weighted average of all current measurements is taken.
For \NE, the prediction from Ref.~\cite{Fleischer:2010ca} is used, which is based on the ratio of branching fractions of the decays \BdDdstK and \BdDstpi
and is updated using their current world averages\cite{PDG2020}.
The measurements and their uncertainties are thus rescaled to take into account the updated external inputs. 
The variation of the \B-meson lifetimes could affect the estimates of the efficiencies used to determine \fsfd;
it has been checked that this effect is negligible compared to the systematic uncertainties 
associated with each measurement.

\begin{table}[tbp]
    \begin{center}
    \caption{External inputs used in the hadronic and semileptonic analyses updated with respect to previous publications.
	    The value of \NE is updated using Ref.~\cite{PDG2020}.
      The values of CKM matrix elements ratio $|V_{us}| / |V_{ud}|$ and of the meson decay constants' ratio $f_K/f_\pi$
     are the same as in Ref.~\cite{LHCb-PAPER-2012-037}.
    }
    \label{tab:external_inputs}

    \scalebox{0.85}{
    \begin{tabular}{l | c c c c  c }
    \toprule
	    Input           & Value & Reference \\
    \midrule
	    $\BF(\DuKpi)$   & $(3.999\pm 0.045) \% $ & \cite{HFLAV18}\\
	    $\BF(\DdKpipi)$ & $(9.38 \pm 0.16) \% $  & \cite{PDG2020}\\
	    $\BF(\DsKKpi)$  & $(5.47 \pm 0.10) \% $  & \cite{HFLAV18,BESIII}\\
    \midrule
	    \tauBs/\tauBd             & $1.006 \pm 0.004$ & \cite{HFLAV18}    \\
	    $(\tauBu+\tauBd)/2\tauBs$ & $1.032 \pm 0.005$ & \cite{HFLAV18} \\
	    $(1-\xi_s)$               & $1.010 \pm 0.005$ & \cite{Bigi_2011} \\
    \midrule
	    \Na           & $1.000 \pm 0.020$ & \cite{Fleischer:2010ca} \\
	    \NF           & $1.000 \pm 0.042$ & \cite{Bordone:2019guc, Bordone:2020gao} \\
	    \NE           & $0.966 \pm 0.062$ &\cite{Fleischer:2010ca,PDG2020} \\
	\midrule 
	    $|\Vus| f_K / |\Vud| f_\pi$ & 0.2767 &\cite{LHCb-PAPER-2012-037}   \\
        \bottomrule
    \end{tabular}}
    \end{center}
\end{table}

\section{Combined fit}
\label{sec:Combined_fit}
The fit to the data is performed as a minimisation of the \chisquare function
\begin{equation}
\label{eq:chisquare}
\chisquare =  ( f(x|\theta) - y ) V^{-1} (f(x|\theta) -y )^T + \sum_i \left(\frac{\theta_i - \hat \theta_i }{\sigma_{\theta_i}} \right)^2 \quad,
\end{equation}
where $f$ is the function describing \fsfd in the data, with $x = \pt$ or $\eta$, and $y$ is the vector containing the central values of the measured observables sensitive to \fsfd, and $V$ is their covariance matrix. The set of parameters to be determined, $\theta$, 
includes a subset of parameters that are constrained to external measurements $\hat \theta_i$ with their uncertainties $\sigma_{\theta_i}$. 
While the first term in Eq.~\ref{eq:chisquare} is due to the experimental data compared with the function to be fitted, 
the second is due to external constraints on some of the parameters. 
These constraints are of two kinds: external constraints on theoretical input parameters 
and overall scaling parameters to take into account scale-related systematic uncertainties for some of the analyses.
These uncertainties are not included in the data points, to avoid the bias described in Ref.~\cite{DAgostini:1993arp}, 
due to the failure of the intrinsic assumptions of the $\chi^2$ method, and are thus taken into account as suggested in Ref.~\cite{Mo:2007aea}. 

The scale factors related to the theoretical inputs, owing to their larger uncertainties, 
are found to have fitted values that differ from the input ones by up to one standard deviation. 
For this reason, these are kept indicated explicitly as ratios of the fitted value to the input value in the presentation of results. 
They are indicated by $r_{AF} = (\Na\NF)^{\rm{fitted}}/(\Na\NF)^{\rm{input}}$ for those common to the hadronic measurements
and as $r_{E} = \NE^{\rm{fitted}}/\NE^{\rm{input}}$ for the exchange-diagram inputs.

The uncertainties from inputs common to the semileptonic and hadronic measurements, 
including the \B-meson lifetimes and \D-meson branching fractions, 
are 100\% correlated among the hadronic measurements and 68\% correlated with the semileptonic measurement,
based on the relative rates of the \BsDsXmunu and \BsDKXmunu decays and of the \BorDuXmunu and \BorDdXmunu decays.

The fit model as a function of \pt assumes the common functional form
\begin{equation}
\label{eq:func}
\fsfdfrac \,(\pt, \sqs) = a + b \cdot \pt \quad . 
\end{equation}
The dependence on collision energy is expressed by letting intercept $a$ and slope $b$ parameters have different values at different~\sqs.
Fits with different functional forms have been performed and the data can also be described with an
exponential, Gaussian, or power-law functions, with similar fit quality.
Attempts to describe the data with other functional forms suggested in Ref.~\cite{Berezhnoy:2015lta} 
resulted in significantly worse fit quality. 
No attempt was made to describe the data with more parameters, with the exception of the physics-motivated fit with the Tsallis-statistics-inspired function, described at the end of the paper. 

The parameters of the default fit are summarised in Table~\ref{tab:parameters} together with the observables to which they are sensitive.
In addition to the $a$ and $b$ parameters of Eq.~\ref{eq:func}, 
the only free parameter is \FR, the ratio of \bsjpsiphi and \bujpsik branching fractions. 
The other parameters are all Gaussian constrained to unity with the relevant uncertainty. They include $r_{AF}$, $r_{E}$ as defined above, $S_1$, the parameter propagating the correlated systematic uncertainty of semileptonic and hadronic measurements due to external parameters, and $S_2$, $S_3$, and $S_4$, the parameters propagating experimental systematic uncertainties on the input measurements. 

\begin{table}[tbp]
    \begin{center}
    \caption{Observables and related parameters of the default fit. See text for a detailed explanation. }.
    \label{tab:parameters}
    \begin{tabular}{l | c c c c  c }
    \toprule
    Observable   & Parameters  &    Fit mode \\ 
    \midrule 
\multirow{2}{*}{\fsfd} &  $a(7\tev)$, $a(8\tev)$, $a(13\tev)$ & Free \\
  & $b(7\tev)$, $b(8\tev)$, $b(13\tev)$ & Free \\
\multirow{2}{*}{$\mathcal{B}(\bsdspi)$} &  $r_{AF}$     &   Gaussian constrained \\ 
 & $r_{E}$     &   Gaussian constrained \\ 
 $\mathcal{B}(\bsjpsiphi)$ & \FR  &     Free \\
 & $S_1$    & Gaussian constrained \\
 &$S_2$, $S_3$, $S_4$ &   Gaussian constrained \\ 
    
        \bottomrule
    \end{tabular}
    \end{center}
\end{table}

\section{Results}
\label{sec:Results}

Results of the default fit are presented in the following described separately for the differential \fsfd results (Sect.~\ref{sec:Results_fsfd}),
for the \bsjpsiphi and \bsdspi branching fractions 
(Sect.~\ref{sec:Results_BF}),
and for the integrated \fsfd 
(Sect.~\ref{sec:Results_fsfd_integrated}). 
Values and uncertainties of the parameters and their correlations are reported in the Supplemental Material~\cite{supplemental}. 

\subsection{Determination of \fsfd}
\label{sec:Results_fsfd}

\begin{figure}[tbp]
     \includegraphics[width =0.32\textwidth ]{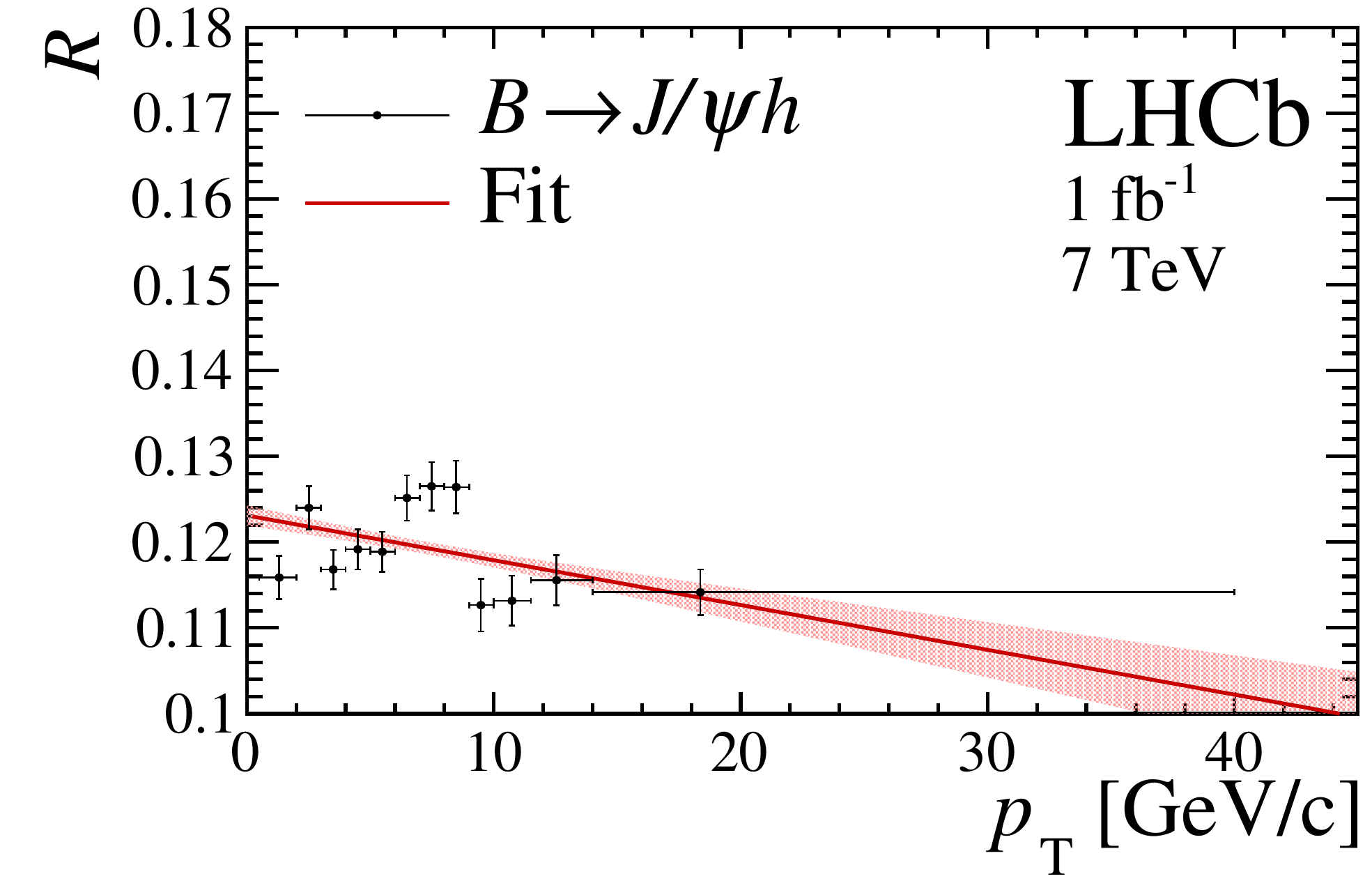}
     \includegraphics[width =0.32\textwidth ]{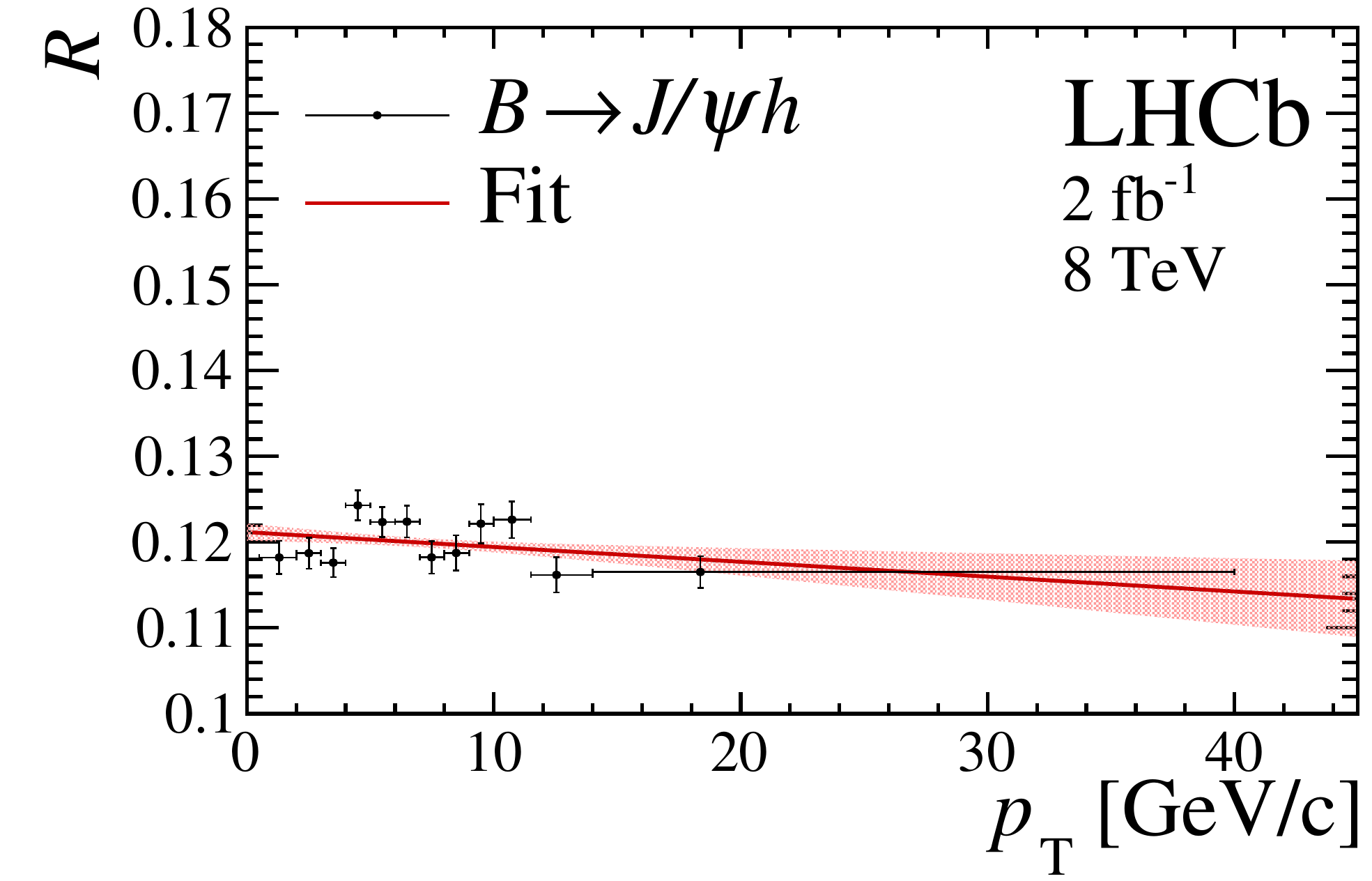}
     \includegraphics[width =0.32\textwidth ]{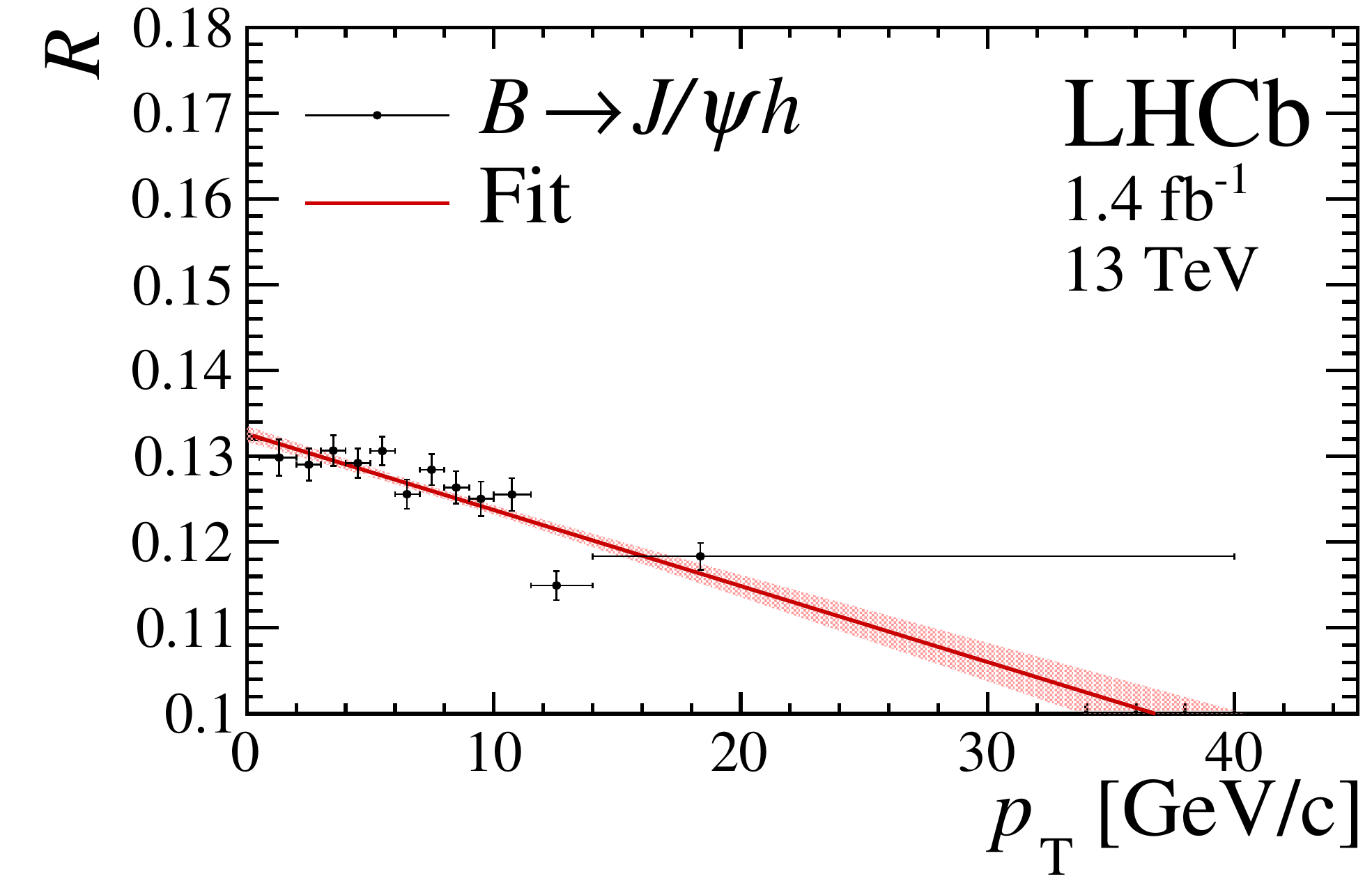}\\
     \includegraphics[width =0.32\textwidth ]{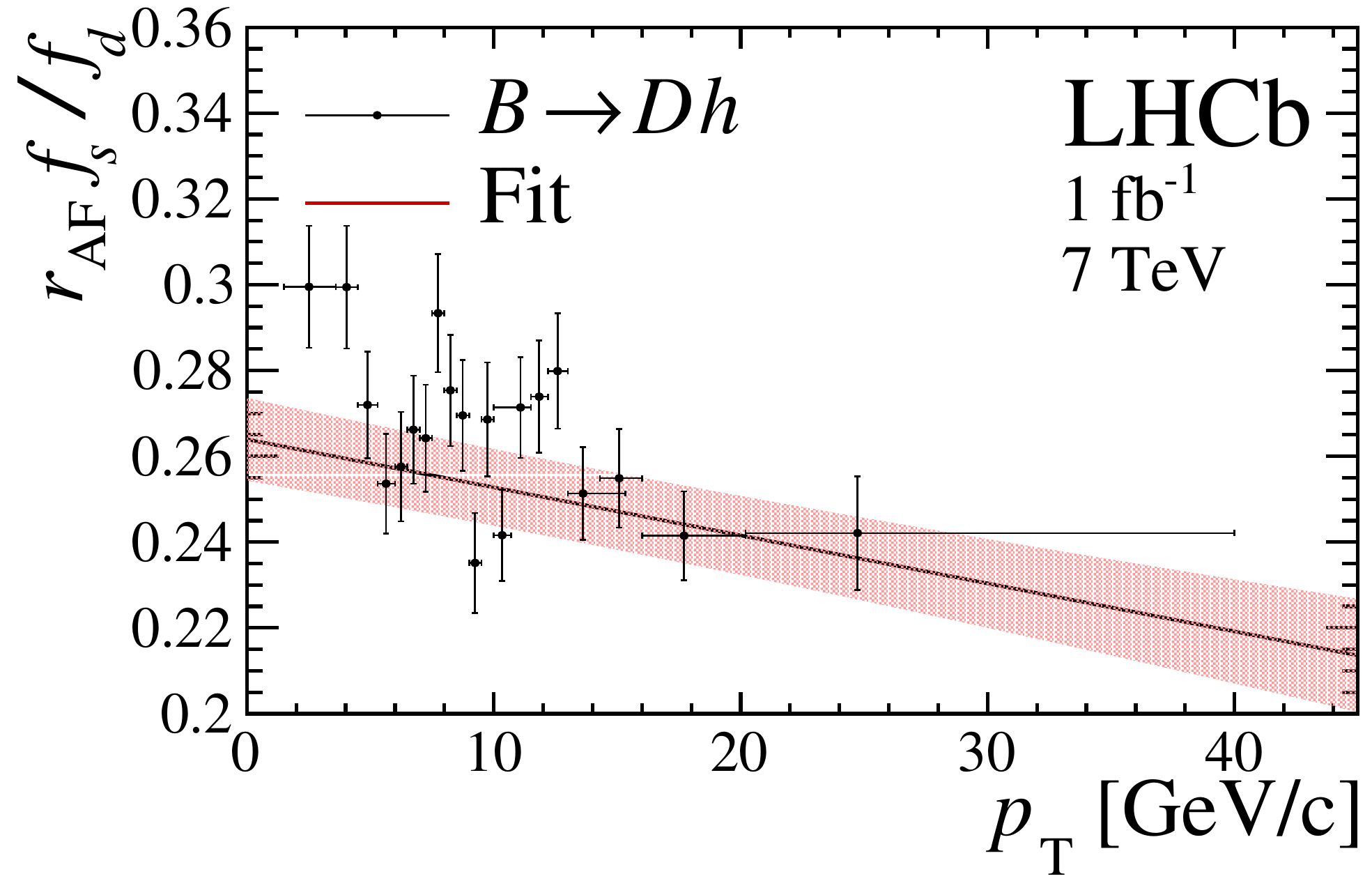}
     \includegraphics[width =0.32\textwidth ]{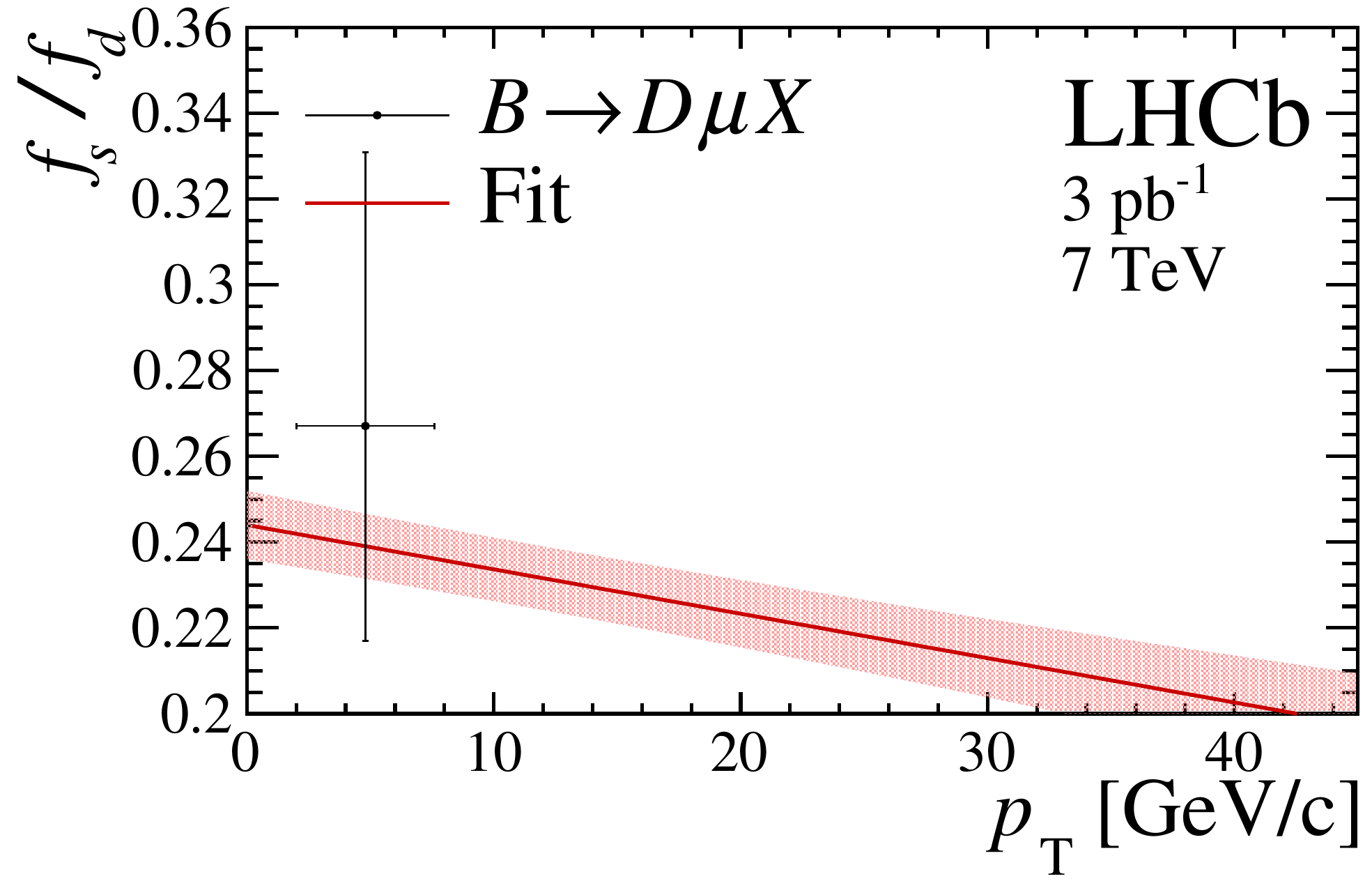}
     \includegraphics[width =0.32\textwidth ]{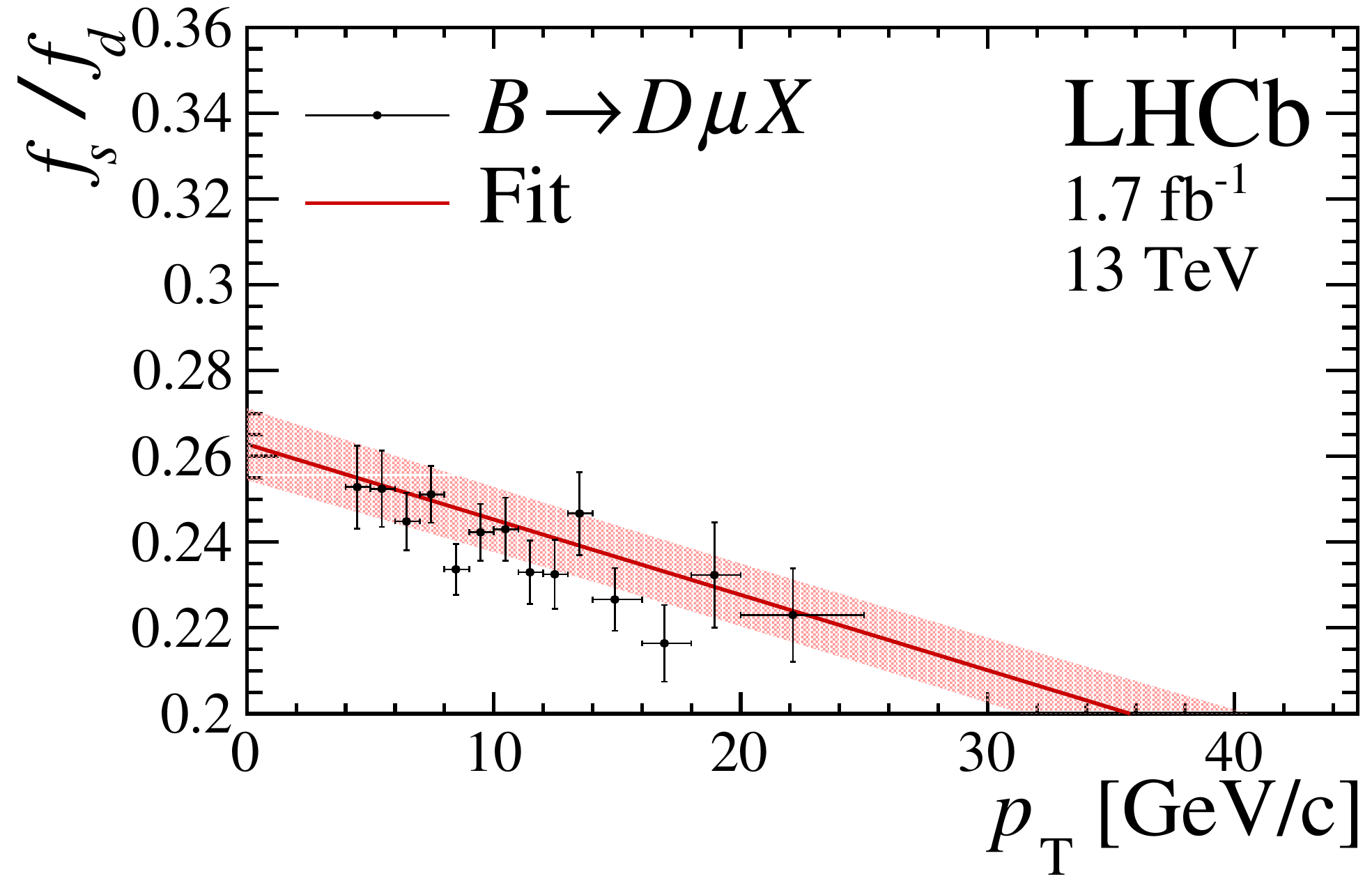}\\
     \includegraphics[width =0.32\textwidth ]{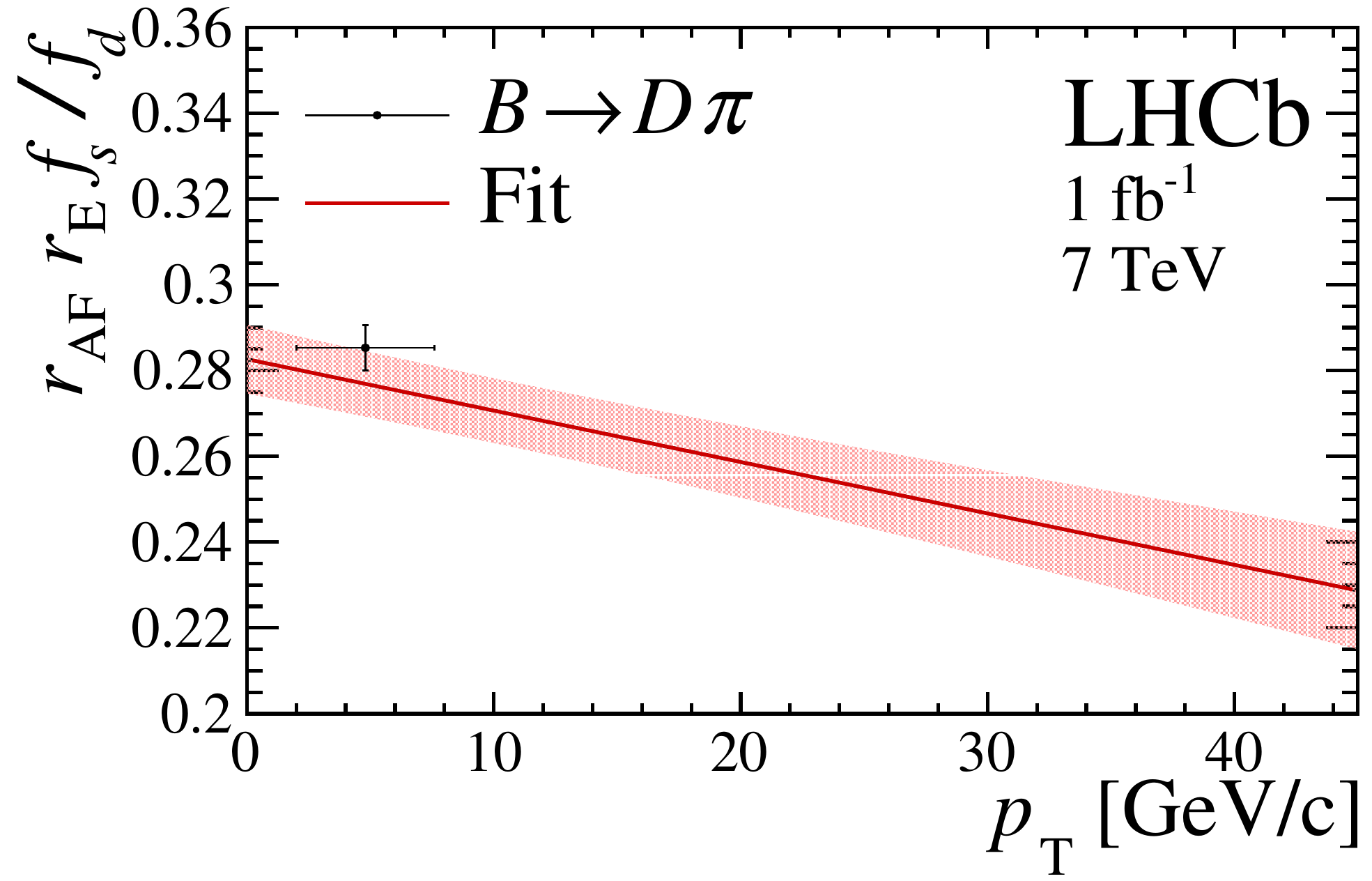}
     \includegraphics[width =0.32\textwidth ]{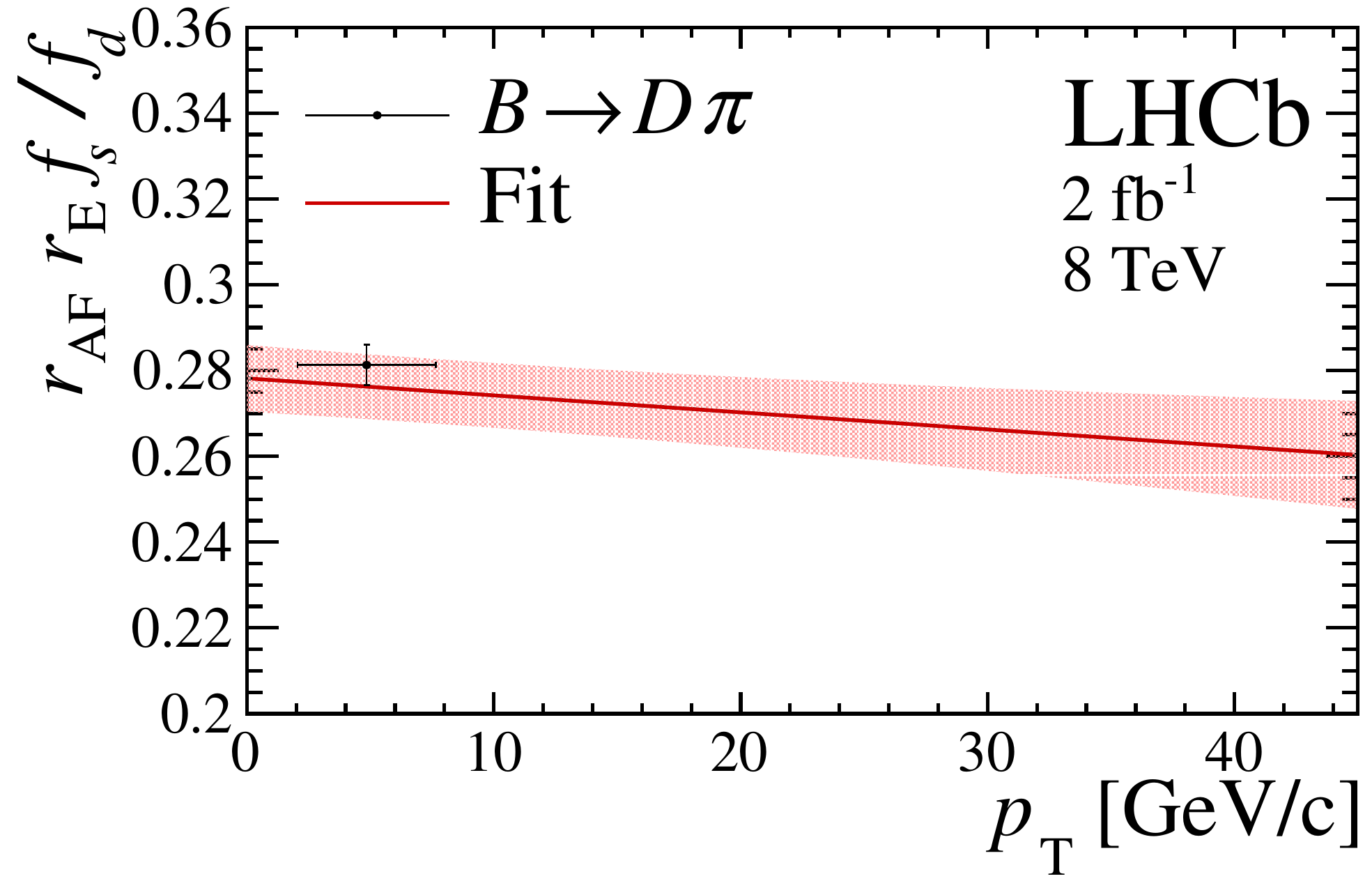}
     \includegraphics[width =0.32\textwidth ]{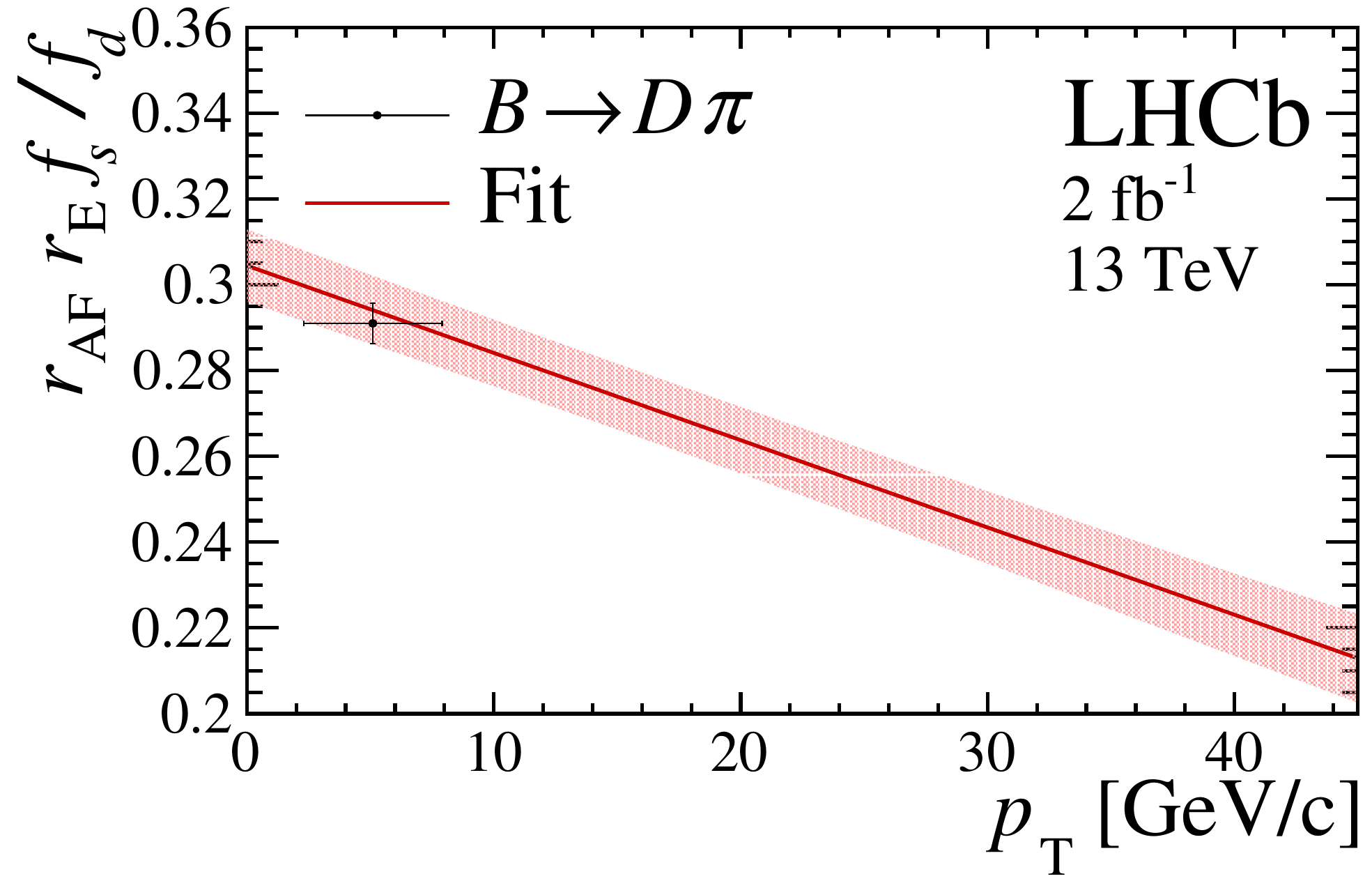}
 \caption{ Measurements of \fsfd sensitive observables as a function of the \B-meson transverse momentum, 
	   \pt, overlaid with the fit function. The scaling factors $r_{AF}$ and $r_{E}$ are defined in the text;
	   the variable \R is defined in Eq.~\ref{eq:R}.  The vertical axes are zero-suppressed.
	   The uncertainties on the data points are fully independent of each other;
	   overall uncertainties for measurements in multiple \pt intervals are 
	   propagated via scaling parameters, as described in the text.
           The band associated with the fit function shows the uncertainty
	   on the post-fit function for each sample.
	   }\label{fig:results}
\end{figure}

The data as a function of \pt together with the result of the fit are shown in Fig.~\ref{fig:results}. 
The obtained functions at the three different energies are
\begin{eqnarray*}
    \fsfd&(\pt, 7\tev) &=  \fsfdseventev ~,\\
    \fsfd&(\pt, 8\tev) &=   \fsfdeighttev ~,\\
    \fsfd&(\pt, 13\tev) &=  \fsfdthirteentev ~,
\end{eqnarray*}
where the \pt is in units of \gevc and the slope parameters are expressed in $(\!\gevc)^{-1}$.
The resulting \chisquare is 133, for a number of effective degrees of freedom of 74. 
The statistical robustness of the procedure has been verified using ensembles of pseudoexperiments. 
They demonstrate that the procedure obtains the correct coverage and minimal bias for the parameters of interest. In the most extreme case, the bias corresponds to about 10\% 
of the uncertainties on the parameters related to the overall scale. 
This is considered negligible and not corrected for.
The p-value of the fit to data, calculated from the distribution of pseudoexperiment \chisq values, is $1.4\times 10^{-4}$.
When artificially increasing the data uncertainties such that the \chisquare corresponds to a p-value of 0.5, 
following similar procedures to those in Ref.~\cite{PDG2020}, the central values and uncertainties obtained in this paper are unchanged,
with the exception of uncertainties on the slopes versus \pt, which would increase by approximately a relative 25\%  but not affect the integrated measurement of \fsfd. More data will be needed to resolve the exact \pt dependence of \fsfd.

Requiring identical intercepts and slopes at the three energies results in significantly worse fit quality, 
with a difference in \chisq of 115 for two fewer parameters. An F-test~\cite{snecdecor1991statistical} is performed to verify 
the significance of the dependence of the intercept on the energy; the difference in \chisquare corresponds to an F-test statistic 
of 13.2 and to a significance of $5.9$ standard deviations ($\sigma$). 
Similarly, but less significantly, requiring only the slope parameters to be common among the energies increases 
the \chisq by 22 for two fewer parameters, corresponding to an F-test significance of $2.7\,\sigma$.

Many of the input measurements also provide results as a function of pseudorapidity, none of them reporting any dependence on $\eta$.
A combined fit as a function of $\eta$ is also performed here. No dependence on pseudorapidity is found
and the \fsfd value is found to be in agreement with the one obtained through the fit as a function of transverse momentum.

\subsection{\bsjpsiphi and \bsdspi branching fractions}
\label{sec:Results_BF}
\begin{figure}[tp]
\begin{center}
  \includegraphics[width = 0.49\textwidth ]{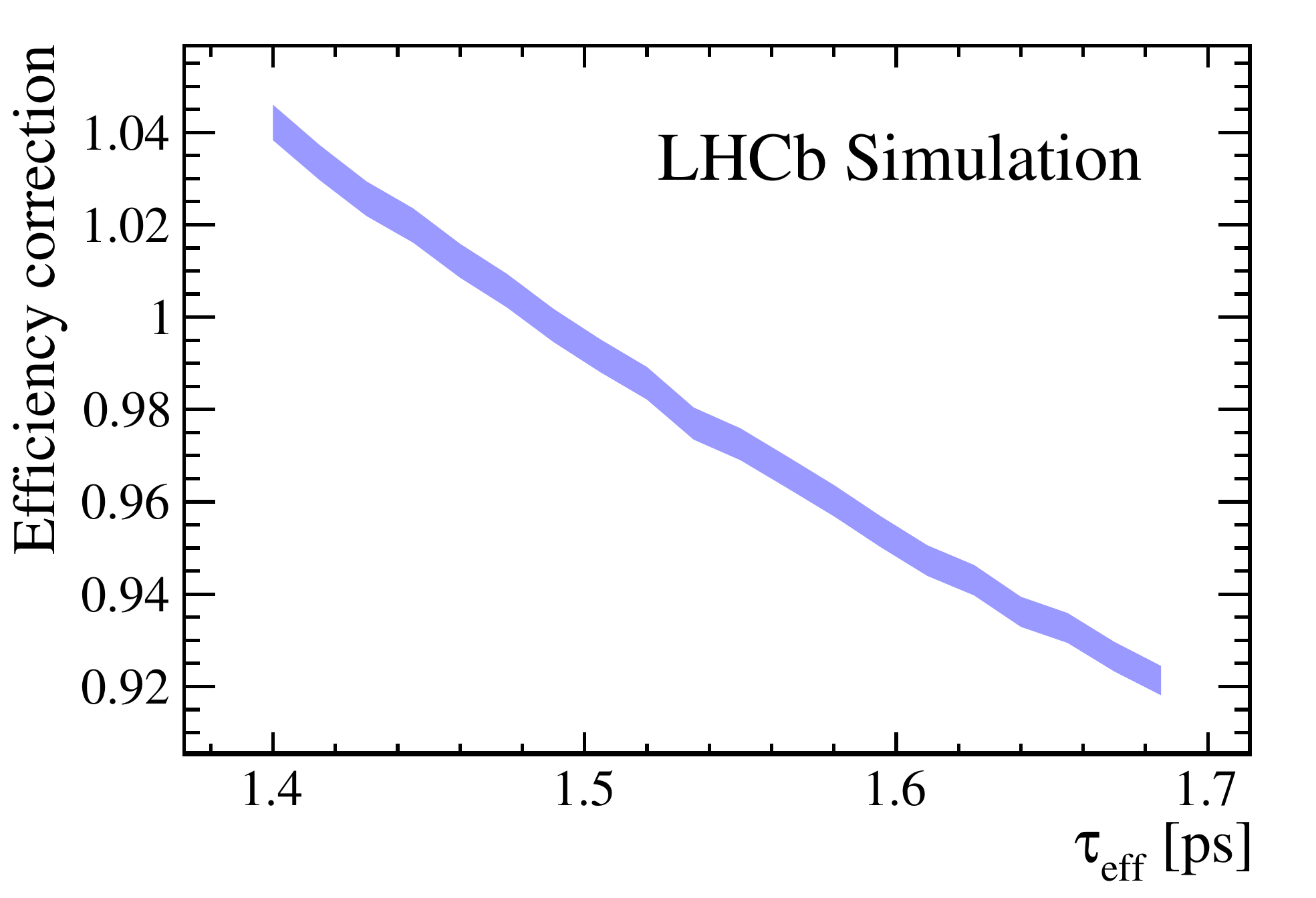}
  \end{center}
 \caption{Efficiency correction versus effective lifetime hypothesis for the \bsjpsiphi branching fraction.
          The band shows the uncertainty on the correction due to the simulated sample size for a given effective lifetime.
	}\label{fig:bsjpsiphi_eff}
\end{figure}

An additional output from the fit is \FR, the ratio of the relative \bsjpsiphi (with $\phi \to \Kp \Km$) to \bujpsik branching fractions, as in Eq.~\ref{eq:R}.
The measurement of the \bsjpsiphi branching fraction reported here is time-integrated, 
and as such should be compared with theoretical predictions that include a correction for the finite \Bs-\Bsb width difference~\cite{DeBruyn:2012wj}. 
In addition, the total efficiency varies for different effective lifetimes;
therefore, branching fraction measurements should be reported for a given effective lifetime value~\cite{Dettori:2018bwt}.
In this paper the results are obtained assuming the \bsjpsiphi parameters measured in Ref.~\cite{LHCb-PAPER-2019-013}, 
which reports the time-dependent analysis of this decay, and the combination with previous LHCb measurements. 
The parameters used in this analysis correspond to a \bsjpsiphi effective lifetime of $\tau_{\rm{eff}} = \taueffbsjpsiphi\ps$, which
 is different by 2.4\% from that used in the simulation for the efficiency in Ref~\cite{LHCb-PAPER-2019-020}. The \R measurements are corrected to take this into account. 
A scaling for different effective lifetimes is reported in Fig.~\ref{fig:bsjpsiphi_eff} 
and should be used as multiplicative correction to recompute the \bsjpsiphi branching fraction under different hypotheses.

The fit value for the \FR parameter is $\FRvalue$.
The uncertainty is reduced to 0.012 when fixing external parameters, the remaining portion is dominated 
by the experimental systematic uncertainties on the input measurements. 
The \FR result can be converted to the \bsjpsiphi branching fraction including the $\phi \to \Kp \Km$ decay branching fraction, 
by multiplying with the \bujpsik branching fraction. The relative production fraction of \Bu and \Bd mesons at \bfactories~\cite{Jung:2015yma},  $1.027 \pm 0.037$, is used to correct the input measurements~\cite{PDG2020} and the \bujpsik branching fraction is found to be $(1.003\pm0.035)\times 10^{-3}$, resulting in
\begin{equation*}
 \BF(\BsToJPsiPhiKK) = \BRBsjpsiphikk \quad,
\end{equation*}
where the first uncertainty includes statistical and systematic uncertainties on the yield ratio as well as the uncertainties on external parameters, and the second arises from the external measurement of \mbox{$\mathcal{B}(\bujpsik)$}.
This result is corrected for the presence of the S-wave component and for the effective lifetime, 
as mentioned earlier.
Taking into account the $\Pphi \to \PK^+ \PK^-$ branching fraction, $(49.2\pm0.5)\%$~\cite{PDG2020},  the \bsjpsiphi branching fraction is 
\begin{equation*}
 \BF(\bsjpsiphi) = \BRBsjpsiphi   \quad,
\end{equation*}
where again the first uncertainty includes statistical and systematic uncertainties on the yield ratios as well as the uncertainties on external parameters, and the second is from external inputs.
This result is compatible with and significantly more precise than the 
PDG world average of $(1.08 \pm 0.08 )\times 10^{-3}$~\cite{PDG2020}. 
It should be noted that the PDG average includes a measurement by the LHCb experiment at $7\tev$
that is at least partially correlated with the $7\tev$ data sample used in the \R measurement included in this paper.
The measurement is consistent with the individual measurements by the 
Belle collaboration, \mbox{$(1.25 \pm 0.07 \pm 0.23) \times 10^{-3}$}~\cite{Thorne:2013llu}, and the CDF collaboration, \mbox{$(1.5 \pm 0.5 \pm 0.1) \times 10^{-3}$}~\cite{Abe:1996kc},
although these have larger uncertainties.

The ratio of the branching fractions of \BsDspi and \BdDpi decays is expressed in terms of the theory parameters in Eq.~\ref{eq:fsfd_hadr}.
However, the theory constraints can be removed and the fit can be repeated to estimate this quantity from data. 
The normalisation of the \fsfd function is correspondingly shifted by a relative 2.5\%, which is within the final uncertainties. 
The other parameters are found to be in good agreement. 
The uncertainties on all parameters do not change significantly with respect to the default fit. 
The output of this fit is then converted to a measurement of the abovementioned ratio of branching fractions.
The result is 
\begin{equation*}
\frac{\mathcal{B}(\BsDspi)}{ \mathcal{B}(\BdDpi)}  = \RHBRvalue \quad,
\end{equation*}
where the correlation of the \D-meson branching fractions is considered when calculating this uncertainty.
The uncertainty is reduced to 0.033 when fixing external parameters; the remaining portion is dominated 
by the experimental systematic uncertainties on the input measurements. 
This result can be compared with the ratio measured by the LHCb collaboration using only 2011 data~\cite{LHCb-PAPER-2011-022}, which yields
$\mathcal{B}(\BsDspi)/\mathcal{B}(\BdDpi)  =  1.10 \pm 0.018 \pm 0.033^{\,+\,0.07}_{\,-\,0.08}$, where the uncertainties 
are statistical, systematic and due to \fsfd, and with the current ratio of  PDG averages of $1.19 \pm 0.19$~\cite{PDG2020}. 
This result is in excellent agreement with both and significantly more precise.
The relative production fraction of \Bu and \Bd mesons at the \bfactories~\cite{Jung:2015yma},  $1.027 \pm 0.037$, is used to correct the input measurements for the \BdDpi branching fraction~\cite{PDG2020}; it is found to be $(2.72\pm0.14)\times 10^{-3}$.
Using this value,
the branching fraction of \BsDspi decays is measured to be 
\begin{equation*}
 \mathcal{B}(\BsDspi)  = \BRBsDspivalue\quad,
\end{equation*}
where the first uncertainty is due to the total experimental uncertainties on the yield ratios and the uncertainties from external parameters and the second is due to the \BdDpi branching fraction. 
This result is in agreement with and significantly more precise than the previous LHCb measurement~\cite{LHCb-PAPER-2011-022},
$\mathcal{B}(\BsDspi) = (2.95 \pm 0.05 \pm 0.17^{\,+\,0.18}_{\,-\,0.22})\times 10^{-3}$, 
where the uncertainties are again statistical, systematic and due to \fsfd, 
and the PDG average, $(3.00 \pm 0.23)\times 10^{-3}$, which is dominated by the latter.

\subsection{Integrated \fsfd results}
\label{sec:Results_fsfd_integrated}
\begin{figure}
\begin{center}
  \includegraphics[width = 0.49\textwidth ]{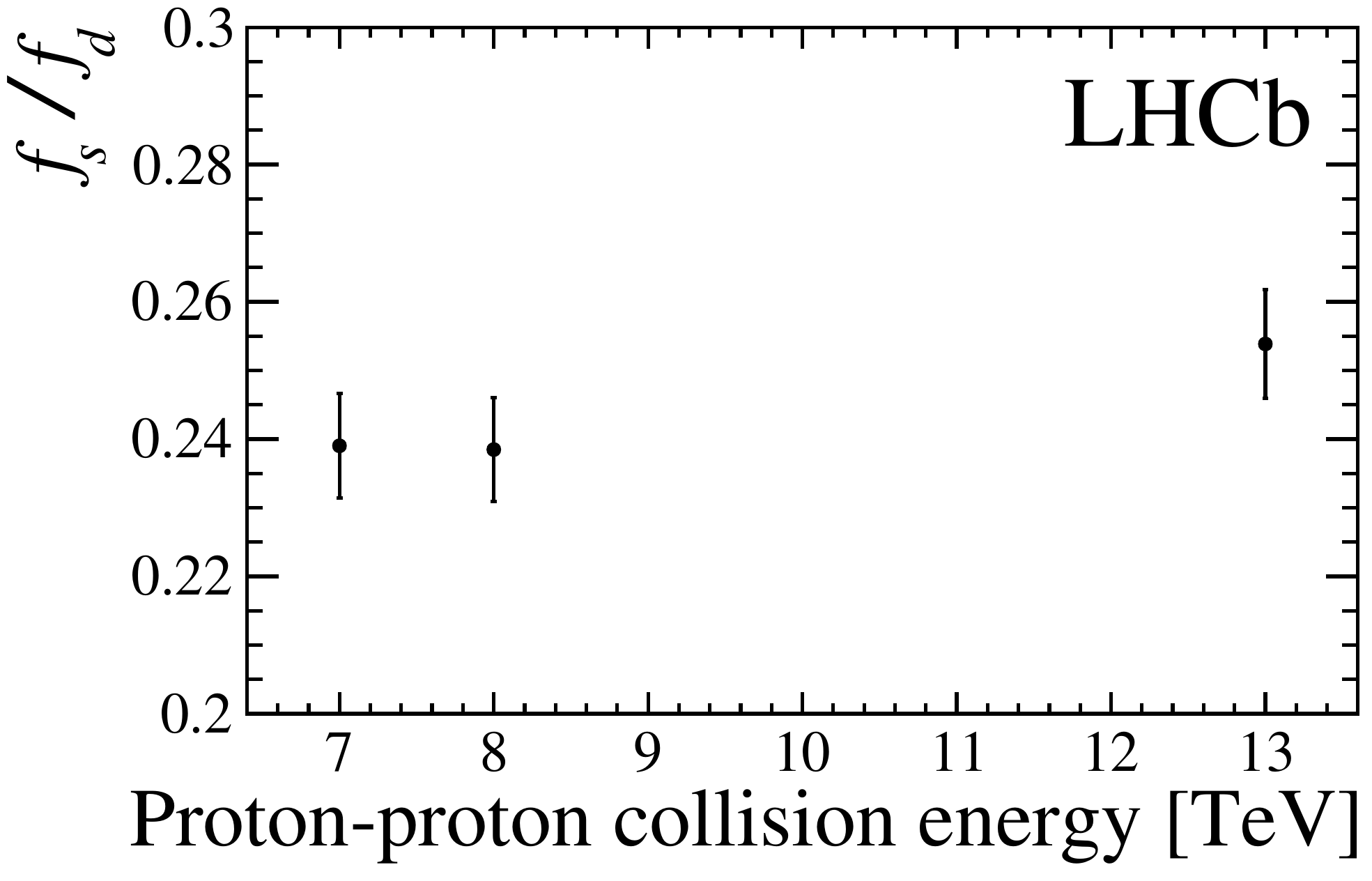}
  \end{center}
 \caption{Fragmentation fraction ratio \fsfd as a function of proton-proton centre-of-mass energy.}\label{fig:fsfdsqs}
\end{figure}

Reference \pt spectra, needed to calculate the integrated \fsfd ratios, are obtained by generating \Bs and \Bd mesons 
in the fiducial acceptance, without any simulation of the detector. 
The average \pt for these spectra are very similar for \Bs and \Bd mesons; they are $4.80$, $4.85$ and $5.10 \gevc$ for the
$\sqs=7$, 8 and $13\tev$ generated samples, respectively, 
with a standard deviation of about $2.8 \gevc$ at all energies.
The following integrated \fsfd values 
for $\pt \in [0.5, 40] \gevc$ and $\eta \in [ 2, 6.4 ]$ are measured
\begin{eqnarray*}
    \label{eq:fsfd_int_results}
    \fsfd\,(7\tev) &=&  \fsfdintseventev ~, \nonumber \\
    \fsfd\,(8\tev) &=&  \fsfdinteighttev ~, \\
    \fsfd\,(13\tev) &=& \fsfdintthirteentev ~, \nonumber
\end{eqnarray*}
which are shown in Fig.~\ref{fig:fsfdsqs}.
Ratios of the integrated values at different energies have also been calculated, incorporating correlations between the uncertainties, yielding 
\begin{eqnarray*}
    \label{eq:fsfd_int_results_ratio}
    \frac{\fsfd\,(13\tev)}{\fsfd\,(7\tev)} &=&  \text{$1.064 \pm 0.008$
 } ~,\\
     \frac{\fsfd\,(13\tev)}{\fsfd\,(8\tev)} &=&  \text{$1.065 \pm 0.007$
} ~,\\
     \frac{\fsfd\,(8\tev)}{\fsfd\,(7\tev)} &=&  \text{$0.998 \pm 0.008$
} ~,  
\end{eqnarray*}
which can be used to correctly normalise future analyses using data at different energies. 
These values are calculated assuming an equal average \pt of 5 \gevc for the different energies, however, it has been verified that varying this assumption 
does not modify the results significantly.
In addition, the ratio of the Run~2 ($13\tev$) over Run 1 (7 and $8\tev$) measurements has been computed, 
weighting the Run~1 values by their integrated luminosity (1 and $2\invfb$, respectively), resulting in 
\begin{equation*}
 \frac{\fsfd\,(\rm{Run~2})}{ \fsfd\,({\rm Run~1})} =  \text{$1.064 \pm 0.007$ } ~.
\end{equation*}

\section{Updated branching fractions measurements}
\label{sec:Updated_BFs}
Using the results for the integrated \fsfd, $\BF(\bsjpsiphi)$ and $\BF(\BsDspi)$, 
previous LHCb measurements of \Bs branching fractions are updated
by scaling these with either \fsfd and a \Bd or \Bu branching fraction, or with the associated normalisation \Bs branching fraction. 
The \Bd and \Bu normalisation branching fractions are updated using the current PDG world averages~\cite{PDG2020}, corrected for the relative production fraction of \Bu and \Bd mesons at the \bfactories~\cite{Jung:2015yma}.
The sole exception is $\BF(\bdjpsikstar)$, for which the branching fraction is taken from the result of the only amplitude analysis, as performed by the Belle experiment~\cite{Chilikin:2014bkk}. 
The \Bd and \Bu normalisation branching fractions are presented in Table~\ref{tab:norm_bfs}.
For \lhcb measurements using both Run~1 and Run~2 data, 
an average \fsfd is estimated using the relative expected yields at the different energies, 
with the uncertainties from \fsfd and normalisation mode branching fractions recomputed accordingly.
Updating these inputs significantly reduces the systematic uncertainty from \fsfd on all previous \Bs branching fraction measurements,
such that the updated results supersede those from the cited publications.
The only exception is the branching fraction of \bsmumu decays,
for which the LHCb result updated here has less precision than the LHC-wide average determined recently~\cite{LHCb-CONF-2020-002}, and which will be superseded only by future updates of this measurement with the full Run~2 data sample.
The updated branching fractions are grouped according to decay type:
rare \Bs decays are updated in Table~\ref{tab:updated_bfs_rare}, \Bs decays with charmonium in Table~\ref{tab:updated_bfs_b2cc},
charmless \Bs decays in Table~\ref{tab:updated_bfs_bnoc}, and \Bs decays with charm mesons in Table~\ref{tab:updated_bfs_b2oc}.
As the estimated value of \fsfd for the Run 1 data samples decreased, in general the values of the branching fractions increase with respect to their published values; the branching fractions normalised to \bsjpsiphi and \bsdspi instead decrease with respect to their published values.

The recent measurement of $|\Vcb|$ with \BsDsorDsstmunu decays using Run 1 data~\cite{LHCb-PAPER-2019-041}, also relies on an estimate of \fsfd and is independent of the uncertainty on the product
$\BF(\DsKKpi)\times \tau_\Bs$.
For this estimate, the correlation of \fsfd with $\BF(\DsKKpi)$ from the semileptonic measurement is used.
The resulting estimates for $|\Vcb|$ are \mbox{$|\Vcb|_{\rm CLN} =$  \text{$ (40.8  \pm 0.6 \pm 0.9 \pm 1.1) \times 10^{-3} $ }}, 
\mbox{$|\Vcb|_{\rm BGL} = $\text{$ (41.7  \pm 0.8 \pm 0.9 \pm 1.1) \times 10^{-3} $ }}, where CLN~\cite{CLN} and BGL~\cite{BGL} stand for two hadronic form-factor parametrisations. 
Both results are consistent with the current world average (see for example Ref.~\cite{PDG2020}).

\begin{table}[tp]
    \begin{center}
    \caption{The branching fractions of \Bd and \Bu normalisation channel decays used to update previous measurements of \Bs branching fractions, as reported in Ref.~\cite{PDG2020} for all but the \mbox{\bdjpsikstar} branching fraction, which is taken from the amplitude analysis in Ref~\cite{Chilikin:2014bkk}, and corrected for the relative production fraction of \Bu and \Bd mesons at \bfactories~\cite{Jung:2015yma}.}
                \label{tab:norm_bfs}
   \begin{tabular}[t]{llllll}
      \toprule
	   Decay mode               & Branching fraction  \\
       \midrule
	   $\bdjpsikstar$&  $(1.21 \pm 0.08) \times 10^{-3}$ \\
	   $\Bd\to\jpsi\rho^0$ &   $(2.58 \pm 0.18) \times 10^{-5}$  \\
	   $\Bd\to\jpsi\KS$ &  $(4.40 \pm 0.17) \times 10^{-3}$  \\
	   $\Bd\to\jpsi\KS\pip\pim$ &  $(2.18 \pm 0.19) \times 10^{-3}$  \\
	   $\Bd\to\psitwos\Kstarz$&  $(5.98 \pm 0.42) \times 10^{-4}$ \\
	   $\Bd\to\psitwos\Kp\pim$&  $(5.88 \pm 0.42) \times 10^{-4}$  \\
       \midrule
	   $\Bd\to\Kp\pim$ &  $(1.98 \pm 0.07) \times 10^{-5}$  \\
	   $\Bd\to\KS\pip\pim$ &  $(2.51 \pm 0.11) \times 10^{-5}$ \\
	   $\Bd\to\Kstarp\pim$ &  $(7.60 \pm 0.43) \times 10^{-6}$  \\
	   $\Bd\to\proton\antiproton\Kp\pim$ &  $(6.30 \pm 0.50) \times 10^{-6}$\\
	   $\Bd\to\proton\Lbar\pim$ &  $(3.18 \pm 0.30) \times 10^{-6}$\\
	   $\Bd\to\Kstarz\gamma$ &  $(4.13 \pm 0.26) \times 10^{-5}$  \\
	   $\Bd\to\phi\KS$ &  $(3.70 \pm 0.36) \times 10^{-6}$  \\
	   $\Bd\to\phi\Kstarz$ &  $(1.01 \pm 0.05) \times 10^{-5}$  \\
	   \bottomrule
	   \end{tabular}
	   \begin{tabular}[t]{ll}
	    \toprule
	   Decay mode               & Branching fraction  \\
	   \midrule
	   $\Bd\to\Dm\mup\neum$ &  $(2.31 \pm 0.10)\%$ \\
	   $\Bd\to\Dstarm\mup\neum$ &  $(5.05 \pm 0.14)\%$  \\
	   $\Bd\to\Dstarpm\Dmp$ &  $(6.2 \pm 0.6) \times 10^{-4}$  \\
	   $\Bd\to\Dp\Dm$ &  $(2.14 \pm 0.19) \times 10^{-4}$  \\
	   $\Bd\to\Dm\Dsp$ &     $(7.3 \pm 0.8) \times 10^{-3}$  \\
	   $\Bu\to\Dzb\Ds$ &     $(9.0 \pm 0.9) \times 10^{-3}$ \\
	   $\Bd\to\Dzb\pip\pim$ & $(8.8 \pm 0.5) \times 10^{-4}$  \\
	   $\Bd\to\Dzb\rho$ & $(3.21 \pm 0.21) \times 10^{-4}$  \\
	   $\Bd\to\Dzb\KS$ &  $(5.3 \pm 0.7) \times 10^{-5}$  \\
	   $\Bd\to\Dzb\Kp\Km$ &  $(6.1 \pm 0.6) \times 10^{-5}$  \\
      \bottomrule
 \end{tabular}
 \end{center}
\end{table}

\begin{table}[t]
    \begin{center}
    \caption{Updated branching fractions of rare \Bs decays. 
             The uncertainties are statistical, systematic,
             due to \fsfd, and due to the normalisation branching fraction.
             The $\Bs\to\phi\mumu$ branching fractions in different $q^2$ intervals, where $q^2$ is defined as dimuon invariant mass squared in \gevcc, are normalised with respect to \bsjpsiphi.
	     Results with the \normcheck symbol have had their normalisation branching fraction updated as well.}
                \label{tab:updated_bfs_rare}
	\begin{adjustwidth}{-0.65cm}{}
	    \scalebox{0.8}{
   \begin{tabular}{llllll}
      \toprule
	   Decay mode               & Updated branching fraction & Previous result &  &  \\
      \midrule
	   $\Bs\to\phi\gamma$&  $ (3.75  \pm 0.18 \pm 0.12 \pm 0.12 \pm 0.24) \times 10^{-5} $  & $(3.52 \pm 0.17 \pm 0.11 \pm 0.29 \pm 0.12) \times 10^{-5}$   & \cite{LHCb-PAPER-2012-019}& \normcheck\\
	   $\Bs\to \mumu$         &   $ (3.26  \pm 0.65  ^{ + 0.22 }_{ - 0.11 }  \pm 0.10) \times 10^{-9} $  & $(3.0 \pm 0.6 ^{+0.2}_{-0.1} \pm 0.2) \times 10^{-9}$ & \cite{LHCb-PAPER-2017-001} &  \\
	   $\Bs\to \Kstarzb\mumu$ &   $ (3.09  \pm 1.07 \pm 0.21 \pm 0.10 \pm 0.22) \times 10^{-8} $ & $(2.9 \pm 1.0 \pm 0.2 \pm 0.2 \pm 0.2) \times 10^{-8}$ &   \cite{LHCb-PAPER-2018-004} & \\
	   $\Bs\to\pip\pim\mumu$ & $ (8.66  \pm 1.50 \pm 0.47 \pm 0.28 \pm 0.60) \times 10^{-8} $ & $(8.6 \pm 1.5 \pm 0.5 \pm 0.5 \pm 0.7) \times 10^{-8}$ &   \cite{LHCb-PAPER-2014-063}& \normcheck\\
       \midrule
       \midrule
	   $\Bs\to\phi\mumu$ &$ (7.54   ^{ + 0.43 }_{ - 0.41 }  \pm 0.30 \pm 0.36) \times 10^{-7} $ & $(7.97 ^{+0.45}_{-0.43} \pm 0.32 \pm 0.60) \times 10^{-7}$  & \cite{LHCb-PAPER-2015-023} & \normcheck\\
       \midrule
       $\quad \qsq \in [ 1.0, 6.0]$ & $ (2.44   ^{ + 0.31 }_{ - 0.30 }  \pm 0.07 \pm 0.12) \times 10^{-8} $    & $(2.58 ^{+0.33}_{-0.31} \pm 0.08 \pm 0.19) \times 10^{-8}$ &   \cite{LHCb-PAPER-2015-023}& \normcheck\\
       $\quad \qsq \in [15.0,19.0]$ & $ (3.82   ^{ + 0.38 }_{ - 0.36 }  \pm 0.12 \pm 0.18) \times 10^{-8} $  & $(4.04 ^{+0.39}_{-0.38} \pm 0.13 \pm 0.30) \times 10^{-8}$ &  \cite{LHCb-PAPER-2015-023}& \normcheck\\
\midrule
       $\quad \qsq \in [0.1, 2.0] $ & $ (5.54   ^{ + 0.69 }_{ - 0.65 }  \pm 0.13 \pm 0.27) \times 10^{-8} $  & $(5.85 ^{+0.73}_{-0.69} \pm 0.14 \pm 0.44) \times 10^{-8}$ &  \cite{LHCb-PAPER-2015-023}& \normcheck\\
       $\quad \qsq \in [2.0, 5.0] $ & $ (2.42   ^{ + 0.40 }_{ - 0.38 }  \pm 0.06 \pm 0.12) \times 10^{-8} $  & $(2.56 ^{+0.42}_{-0.39} \pm 0.06 \pm 0.19) \times 10^{-8}$ &  \cite{LHCb-PAPER-2015-023}& \normcheck\\
       $\quad \qsq \in [5.0, 8.0] $ & $ (3.03   ^{ + 0.42 }_{ - 0.40 }  \pm 0.07 \pm 0.15) \times 10^{-8} $  & $(3.21 ^{+0.44}_{-0.42} \pm 0.08 \pm 0.24) \times 10^{-8}$ &  \cite{LHCb-PAPER-2015-023}& \normcheck\\
       $\quad \qsq \in [11.0, 12.5]$ & $ (4.45   ^{ + 0.65 }_{ - 0.62 }  \pm 0.14 \pm 0.21) \times 10^{-8} $  & $(4.71 ^{+0.69}_{-0.65} \pm 0.15 \pm 0.36) \times 10^{-8}$ &  \cite{LHCb-PAPER-2015-023}& \normcheck\\
       $\quad \qsq \in [15.0, 17.0]$ & $ (4.28   ^{ + 0.54 }_{ - 0.51 }  \pm 0.11 \pm 0.21) \times 10^{-8} $  & $(4.52 ^{+0.57}_{-0.54} \pm 0.12 \pm 0.34) \times 10^{-8}$ &  \cite{LHCb-PAPER-2015-023}& \normcheck\\
       $\quad \qsq \in [17.0, 19.0]$ & $ (3.75   ^{ + 0.54 }_{ - 0.51 }  \pm 0.13 \pm 0.18) \times 10^{-8} $  & $(3.96 ^{+0.57}_{-0.54} \pm 0.14 \pm 0.30) \times 10^{-8}$ &  \cite{LHCb-PAPER-2015-023}& \normcheck\\
      \bottomrule
 \end{tabular}}
	\end{adjustwidth}
 \end{center}
\end{table}

\begin{table}[tb]
\begin{center}
    \caption{Updated branching fractions of \Bs decays with charmonia in the final state.
            The uncertainties are statistical, systematic,
             due to \fsfd, and due to  the normalisation branching fraction.
             The second, third and fourth set of branching fractions are normalised to \bsjpsiphi, $\Bs\to\jpsi\eta^{(')}$, \bsjpsipipi, respectively,
             and their third uncertainty covers the full normalisation uncertainty.
             Results with the \normcheck symbol have had their normalisation branching fraction updated as well.
             }
 \label{tab:updated_bfs_b2cc}
	\begin{adjustwidth}{-1cm}{}
 \scalebox{0.8}{
  \begin{tabular}{llllll}
      \toprule
   Decay mode               & Updated branching fraction & Previous result &  &  \\
    \midrule

	  $\Bs\to\jpsi\KS$  &  $ (2.06  \pm 0.08 \pm 0.06 \pm 0.07 \pm 0.08) \times 10^{-5} $ & $(1.93 \pm 0.08 \pm 0.05 \pm 0.11 \pm 0.07) \times 10^{-5}$ & \cite{LHCb-PAPER-2015-005} & \\
	  $\Bs\to\jpsi\KS\Kpm\pimp$ &  $ (5.01  \pm 0.35 \pm 0.33 \pm 0.16 \pm 0.44) \times 10^{-4} $ & $(4.6 \pm 0.3 \pm 0.3 \pm 0.3 \pm 0.4) \times 10^{-4}$ &  \cite{LHCb-PAPER-2014-016} & \normcheck \\ 

	  $\Bs\to\psitwos\Kstarzb$ &  $ (3.62  \pm 0.37 \pm 0.26 \pm 0.12 \pm 0.25) \times 10^{-5} $ &$(3.35 \pm 0.34 \pm 0.24 \pm 0.19 \pm 0.22) \times 10^{-5}$ & \cite{LHCb-PAPER-2015-010} & \\
	  $\Bs\to\psitwos\Kp\pim$&  $ (3.43  \pm 0.23 \pm 0.14 \pm 0.11 \pm 0.24) \times 10^{-5} $   &$(3.12 \pm 0.21 \pm 0.13 \pm 0.18 \pm 0.22) \times 10^{-5}$ & \cite{LHCb-PAPER-2015-010} & \\

	  $\Bs\to\jpsi\eta$        &  $ (4.04  \pm 0.35  ^{ + 0.32 }_{ - 0.43 }  \pm 0.13 \pm 0.28) \times 10^{-4} $  & $(3.79 \pm 0.31 ^{+0.20}_{-0.41} \pm 0.28  \pm 0.56) \times 10^{-4}$ &  \cite{LHCb-PAPER-2012-022}& \normcheck\\
        $\Bs\to\jpsi\etapr$    &  $ (3.67  \pm 0.32  ^{ + 0.14 }_{ - 0.38 }  \pm 0.12 \pm 0.25) \times 10^{-4} $ & $(3.42 \pm 0.30 ^{+0.14}_{-0.35} \pm 0.26 \pm 0.51) \times 10^{-4}$ &  \cite{LHCb-PAPER-2012-022} & \normcheck\\

     \midrule

        $\Bs\to\psitwos\phi$    &  $ (4.98  \pm 0.26 \pm 0.24 \pm 0.24) \times 10^{-4} $  & $(5.33 \pm 0.28 \pm 0.26 ^{+1.37}_{-1.12}) \times 10^{-4}$ &  \cite{LHCb-PAPER-2012-010} & \normcheck \\
        $\Bs\to\chicone\phi$    &  $ (1.92  \pm 0.18 \pm 0.14 \pm 0.09) \times 10^{-5} $ &  $(1.98 \pm 0.19 \pm 0.15 \pm 0.20) \times 10^{-5}$ &  \cite{LHCb-PAPER-2013-024} & \normcheck \\
        $\bsjpsipipi$      &  $ (2.01  \pm 0.05 \pm 0.05 \pm 0.10) \times 10^{-4} $         & $(2.16 \pm 0.05 \pm 0.06 ^{+0.51}_{-0.42}) \times 10^{-4}$ &  \cite{LHCb-PAPER-2012-005}& \normcheck\\
        $\Bs\to\jpsi\phi\phi$   &  $ (1.17  \pm 0.12  ^{ + 0.05 }_{ - 0.09 }  \pm 0.06) \times 10^{-5} $ &  $(1.19 \pm 0.12 ^{+0.05}_{-0.09} \pm 0.10) \times 10^{-5}$ &  \cite{LHCb-PAPER-2015-033}& \normcheck \\
        $\Bs\to\jpsi\Kstarzb$       &  $ (4.12  \pm 0.19 \pm 0.13 \pm 0.20) \times 10^{-5} $ & $(4.20 \pm 0.20 \pm 0.13 \pm 0.36        ) \times 10^{-5}$ &  \cite{LHCb-PAPER-2015-034}& \normcheck \\
        $\Bs\to\jpsi p\bar{p}$  &  $ (3.54  \pm 0.19 \pm 0.24 \pm 0.16) \times 10^{-6} $ & $(3.58 \pm 0.19 \pm 0.24 \pm 0.30) \times 10^{-6}$ &  \cite{LHCb-PAPER-2018-046}& \normcheck\\ 
        $\Bd\to\jpsi p\bar{p}$  &  $ (3.94  \pm 0.35 \pm 0.26 \pm 0.13) \times 10^{-7} $ & $(4.51 \pm 0.40 \pm 0.30 \pm 0.32) \times 10^{-7}$ &  \cite{LHCb-PAPER-2018-046}& \normcheck\\
 
     \midrule
        $\Bs\to\psitwos\eta$       &  $ (3.35  \pm 0.57 \pm 0.48 \pm 0.50) \times 10^{-4} $  & $(3.15 \pm 0.53 \pm 0.45 ^{+0.61}_{-0.67}) \times 10^{-4}$ &  \cite{LHCb-PAPER-2012-053}& \normcheck\\
        $\Bs\to\psitwos\etapr$  &  $ (1.42  \pm 0.33 \pm 0.06 \pm 0.20) \times 10^{-4} $ & $(1.32 \pm 0.31 \pm 0.05 ^{+0.26}_{-0.28}) \times 10^{-4}$ &  \cite{LHCb-PAPER-2014-056} & \normcheck  \\

     \midrule

        $\Bs\to\jpsi \pip\pim\pip\pim$       &  $ (7.49  \pm 0.30 \pm 0.44 \pm 0.42) \times 10^{-5} $ & $(7.62 \pm 0.36 \pm 0.64 \pm 0.42) \times 10^{-5}$ &  \cite{LHCb-PAPER-2013-055} & \normcheck\\
        $\Bs\to\psitwos\pip\pim$  &  $ (6.87  \pm 0.81 \pm 0.65 \pm 0.39) \times 10^{-5} $  & $(7.3 \pm 0.9 \pm 0.6 ^{+1.9}_{-1.6}) \times 10^{-5}$ &  \cite{LHCb-PAPER-2012-053} & \normcheck\\
    \bottomrule

 \end{tabular}}
	\end{adjustwidth}
 \end{center}
\end{table}

    \begin{table}[tb]
        \begin{center}
    \caption{Updated branching fractions of \Bs decays with a charmless final state.
             The uncertainties are statistical, systematic,
             due to \fsfd,  and due to the normalisation branching fraction.
             The last two branching fractions are normalised with respect to \bsphiphi,
             and their third uncertainty covers the full normalisation uncertainty.
             Results with the \normcheck symbol have had their normalisation branching fraction updated as well.
		}
 \label{tab:updated_bfs_bnoc}
	\begin{adjustwidth}{-0.65cm}{}
 \scalebox{0.8}{
  \begin{tabular}{llllll}
      \toprule
	  Decay mode               & Updated branching fraction & Previous result & & \\
      \midrule

	  $\Bs\to\pip\pim $ &  $ (7.60  \pm 0.58 \pm 0.69 \pm 0.25 \pm 0.25) \times 10^{-7} $ & $(6.91 \pm 0.54 \pm 0.63 \pm 0.40 \pm 0.19 ) \times 10^{-7}$ & \cite{LHCb-PAPER-2016-036} & \\
        $\Bs\to\Km\pip $  &  $ (6.15  \pm 0.49 \pm 0.49 \pm 0.20 \pm 0.20) \times 10^{-6} $   & $(5.4  \pm 0.4  \pm 0.4  \pm 0.4  \pm 0.2  ) \times 10^{-6}$   & \cite{LHCb-PAPER-2012-002}&  \normcheck \\
        $\Bs\to\Kp\Km $ &  $ (2.63  \pm 0.08 \pm 0.16 \pm 0.09 \pm 0.09) \times 10^{-5} $     & $(2.30 \pm 0.07 \pm 0.14 \pm 0.17 \pm 0.07 ) \times 10^{-5}$   & \cite{LHCb-PAPER-2012-002}& \normcheck  \\

        \midrule 
	  $\Bs\to\KS \KS $  &  $ (8.28  \pm 1.60 \pm 0.90 \pm 0.26 \pm 0.81) \times 10^{-6} $ & $(8.3 \pm 1.6\pm 0.9\pm0.3\pm 0.8)\times10^{-6}$ & \cite{LHCb-PAPER-2019-030} & \\

	  $\Bs\to\KS\pip\pim$ &  $ (5.21  \pm 0.74 \pm 0.85 \pm 0.17 \pm 0.23) \times 10^{-6} $  & $(4.7 \pm0.7\pm0.8\pm0.3\pm0.2 )\times 10^{-6}$ & \cite{LHCb-PAPER-2017-010} & \\
	  $\Bs\to\KS\Kpm\pimp$   &  $ (4.64  \pm 0.19 \pm 0.30 \pm 0.15 \pm 0.21) \times 10^{-5} $ & $(4.22\pm0.18\pm 0.28\pm0.25\pm 0.17)\times10^{-5}$ & \cite{LHCb-PAPER-2017-010} & \\
        \midrule
        $\Bs\to\Kstarz\Kstarzb$  &  $ (2.70  \pm 0.44 \pm 0.43 \pm 0.09 \pm 0.19) \times 10^{-5} $ & $(2.81\pm0.46\pm 0.43\pm0.34\pm 0.13)\times10^{-5}$  & \cite{LHCb-PAPER-2011-012}& \normcheck\\

	  $\Bs\to\Kstarpm \Kmp$  &  $ (1.23  \pm 0.18 \pm 0.13 \pm 0.04 \pm 0.07) \times 10^{-5} $  & $(1.27 \pm 0.19 \pm 0.13 \pm 0.07 \pm 0.10 ) \times 10^{-5}$ & \cite{LHCb-PAPER-2014-043} & \\
	  $\Bs\to\Kstarm \pip$ &  $ (3.21  \pm 1.07 \pm 0.41 \pm 0.10 \pm 0.18) \times 10^{-6} $ & $(3.3  \pm 1.1  \pm 0.4  \pm 0.2  \pm 0.3  ) \times 10^{-6}$ & \cite{LHCb-PAPER-2014-043} & \\

        \midrule 
	  $\Bs\to p\bar{p}\Kpm\pimp$ &  $ (1.41  \pm 0.23 \pm 0.12 \pm 0.05 \pm 0.11) \times 10^{-6} $ & $(1.30 \pm 0.21 \pm 0.11 \pm 0.09 \pm 0.08   ) \times 10^{-6}$ & \cite{LHCb-PAPER-2017-005} & \\
	  $\Bs\to \porpbar \LorLbar \Kmp$   &  $ (6.01  \pm 0.66 \pm 0.62 \pm 0.20 \pm 0.57) \times 10^{-6} $ & $(5.46 \pm 0.61 \pm 0.57 \pm 0.32 \pm 0.50  ) \times 10^{-6}$ & \cite{LHCb-PAPER-2017-012} &\\
        \midrule
        $\Bs\to\phi\Kstarzb$&      $ (1.27  \pm 0.28 \pm 0.16 \pm 0.04 \pm 0.07) \times 10^{-6} $ & $(1.10 \pm 0.24 \pm 0.13 \pm 0.08 \pm 0.06 ) \times 10^{-6}$   & \cite{LHCb-PAPER-2013-012}& \normcheck\\
	  $\bsphiphi$&      $ (2.02  \pm 0.05 \pm 0.08 \pm 0.07 \pm 0.11) \times 10^{-5} $ & $(1.84 \pm 0.05 \pm 0.07 \pm 0.11 \pm 0.12 ) \times 10^{-5}$ & \cite{LHCb-PAPER-2015-028} &  \\
        \midrule
        $\Bs\to\phi\pip\pim$&    $ (3.82  \pm 0.25 \pm 0.19 \pm 0.30) \times 10^{-6} $ &$(3.48 \pm 0.23 \pm 0.17 \pm 0.35  ) \times 10^{-6}$   & \cite{LHCb-PAPER-2016-028}& \normcheck\\
        $\Bs\to\phi\phi\phi$&  $ (2.36  \pm 0.61 \pm 0.30 \pm 0.19) \times 10^{-6} $ &  $(2.15 \pm 0.54 \pm 0.28 \pm 0.21  ) \times 10^{-6}$   & \cite{LHCb-PAPER-2017-007}& \normcheck\\

      \bottomrule
 \end{tabular}}
	\end{adjustwidth}

 \end{center}
\end{table}

\begin{table}[tb]
     \begin{center}
    \caption{Updated branching fractions of \Bs decays to open-charm final states.
             The uncertainties are statistical, systematic,
             due to \fsfd,  and due to the normalisation branching fraction.    
             The \BsDsK, \BsDspipipi and  $\Bs\to\Dsm \Kp\pim\pip, \Bs\to\Dsonem\pip$   branching fractions are normalised with respect to \BsDspi and \BsDspipipi, respectively,
             and their third uncertainty covers the full normalisation uncertainty.
             Results with the \normcheck symbol have had their normalisation branching fraction updated as well.
	     }
 \label{tab:updated_bfs_b2oc}
	\begin{adjustwidth}{-1cm}{}
 \scalebox{0.8}{
  \begin{tabular}{llllll}
      \toprule
   Decay mode               & Updated branching fraction & Previous result &  &  & \\

      \midrule
	  $\Bs\to\Dssm \mup\neum$&  $ (5.19  \pm 0.24 \pm 0.47 \pm 0.13 \pm 0.14) \times 10^{-2} $ &         $(5.38\pm 0.25\pm 0.48\pm 0.20 \pm 0.15)\times 10^{-2}$ & \cite{LHCb-PAPER-2019-041} & \\
	  $\Bs\to\Dsm  \mup\neum$&  $ (2.40  \pm 0.12 \pm 0.15 \pm 0.06 \pm 0.10) \times 10^{-2} $ & $(2.49\pm 0.12\pm 0.16\pm 0.09 \pm 0.11)\times 10^{-2}$ & \cite{LHCb-PAPER-2019-041} & \\
     \midrule
	  $\Bs\to\Dp\Dsm$&  $ (3.01  \pm 0.32 \pm 0.10 \pm 0.08 \pm 0.34) \times 10^{-4} $ & $(2.7 \pm 0.3 \pm 0.1 \pm 0.2 \pm 0.3 )\times 10^{-4}$ & \cite{LHCb-PAPER-2014-002} & \\

	  $\Bs\to\Dp\Dm$&  $ (2.47  \pm 0.46 \pm 0.23 \pm 0.08 \pm 0.22) \times 10^{-4} $ & $(2.2 \pm 0.4 \pm 0.1 \pm 0.1 \pm 0.3 )\times 10^{-4}$ & \cite{LHCb-PAPER-2012-050} & \\
	  $\Bs\to\Dz\Dzb$&  $ (1.83  \pm 0.29 \pm 0.29 \pm 0.05 \pm 0.18) \times 10^{-4} $ & $(1.9 \pm 0.3 \pm 0.2 \pm 0.2 \pm 0.3 )\times 10^{-4}$ & \cite{LHCb-PAPER-2012-050} & \\
	  $\Bs\to\Dsp\Dsm$ &  $ (4.38  \pm 0.23 \pm 0.31 \pm 0.11 \pm 0.49) \times 10^{-3} $  & $(4.0 \pm 0.2 \pm 0.2 \pm 0.2 \pm 0.4 )\times 10^{-3}$ & \cite{LHCb-PAPER-2012-050} & \\
	  $\Bs\to\Dstarpm\Dstarmp$ &  $ (8.38  \pm 1.02 \pm 0.12 \pm 0.26 \pm 0.81) \times 10^{-5} $
 & $(8.41 \pm 1.02 \pm 0.12 \pm 0.39 \pm 0.79 )\times 10^{-5}$ & \cite{LHCb-PAPER-2020-037} & \\

        \midrule
	  $\Bs\to\Ds^{(*)}\Dsm^{(*)}$ &  $ (3.36  \pm 0.11 \pm 0.14 \pm 0.09 \pm 0.38) \times 10^{-2} $ & $(3.05\pm 0.10\pm 0.13\pm 0.14\pm 0.34)\times 10^{-2}$ & \cite{LHCb-PAPER-2015-053} & \\
	  $\Bs\to\Dsspm\Dsmp$  &  $ (1.49  \pm 0.06 \pm 0.07 \pm 0.04 \pm 0.17) \times 10^{-2} $ &        $(1.35\pm 0.06\pm 0.06\pm 0.06\pm 0.15)\times 10^{-2}$ & \cite{LHCb-PAPER-2015-053} & \\
	  $\Bs\to\Dssp\Dssm$&  $ (1.39  \pm 0.09 \pm 0.10 \pm 0.04 \pm 0.16) \times 10^{-2} $ &         $(1.27\pm 0.08\pm 0.09\pm 0.06\pm 0.14)\times 10^{-2}$ & \cite{LHCb-PAPER-2015-053} & \\
          \midrule        
	  $\Bs\to\Dzb  \KS$ & $ (4.69  \pm 0.51 \pm 0.28 \pm 0.15 \pm 0.64) \times 10^{-4} $   & $(4.3  \pm 0.5  \pm 0.3  \pm 0.3  \pm 0.6 ) \times 10^{-4}$ & \cite{LHCb-PAPER-2015-050} & \\
	  $\Bs\to\Dstarzb \KS$  & $ (3.05  \pm 1.13 \pm 0.40 \pm 0.10 \pm 0.41) \times 10^{-4} $ & $(2.8  \pm 1.0  \pm 0.3  \pm 0.2  \pm 0.4 ) \times 10^{-4}$ & \cite{LHCb-PAPER-2015-050} & \\
       \midrule
        $\Bs\to\Dzb \Kstarzb$ & $ (5.31  \pm 1.22 \pm 0.54 \pm 0.17 \pm 0.35) \times 10^{-4} $
  & $(4.72 \pm 1.07 \pm 0.48 \pm 0.37 \pm 0.74) \times 10^{-4}$ &  \cite{LHCb-PAPER-2011-008} & \normcheck\\
        $\Bs\to\Dzb \Km\pip$ & $ (1.11  \pm 0.05 \pm 0.07 \pm 0.04 \pm 0.06) \times 10^{-3} $  & $(1.00 \pm 0.04 \pm 0.06 \pm 0.08 \pm 0.10) \times 10^{-3}$ &  \cite{LHCb-PAPER-2013-022} & \normcheck\\
        $\Bs\to\Dzb \phi$ & $ (3.25  \pm 0.38 \pm 0.19 \pm 0.11 \pm 0.18) \times 10^{-5} $   & $(3.0  \pm 0.3 \pm 0.2  \pm 0.2 \pm 0.2) \times 10^{-5}$ &  \cite{LHCb-PAPER-2018-015} & \normcheck\\
        $\Bs\to\Dstarzb \phi$ & $ (4.01  \pm 0.48 \pm 0.27 \pm 0.13 \pm 0.23) \times 10^{-5} $ & $(3.7  \pm 0.5 \pm 0.2  \pm 0.2 \pm 0.2) \times 10^{-5}$ &  \cite{LHCb-PAPER-2018-015}& \normcheck\\
        $\Bs\to\Dzb\Kp\Km$& $ (6.13  \pm 0.59 \pm 0.28 \pm 0.20 \pm 0.56) \times 10^{-5} $ & $(5.7  \pm 0.5 \pm 0.2  \pm 0.3 \pm 0.5) \times 10^{-5}$ &  \cite{LHCb-PAPER-2018-014}& \normcheck\\
    \midrule
        $\BsDsK$&  $ (2.41  \pm 0.05 \pm 0.06 \pm 0.14) \times 10^{-4} $ &         $(2.29\pm 0.05\pm 0.06\pm 0.17)\times 10^{-4}$ &  \cite{LHCb-PAPER-2014-064}& \normcheck\\
        $\BsDspipipi$&  $ (6.43  \pm 1.18 \pm 0.64 \pm 0.38) \times 10^{-3} $ &         $(6.01\pm 1.11\pm 0.60\pm 0.48)\times 10^{-3}$ &  \cite{LHCb-PAPER-2011-016}& \normcheck\\
     \midrule
	  $\Bs\to\Dsm \Kp\pim\pip$&  $ (3.34  \pm 0.32 \pm 0.19 \pm 0.73) \times 10^{-4} $ &         $(3.13\pm 0.30\pm 0.18\pm 0.76)\times 10^{-4}$ &  \cite{LHCb-PAPER-2012-033}& \normcheck\\
        $\Bs\to\Dsonem\pip$&  $ (2.57  \pm 0.64 \pm 0.26 \pm 0.56) \times 10^{-5} $ & $(2.41\pm 0.60\pm 0.24\pm 0.58)\times 10^{-5}$ &  \cite{LHCb-PAPER-2012-033}& \normcheck\\
      \bottomrule
 \end{tabular}}
	\end{adjustwidth}
 \end{center}
\end{table}

\section{Fit to \fsfd with a Tsallis function}
\label{sec:Tsallis}
\begin{figure}[!htbp]
     \includegraphics[width =0.32\textwidth ]{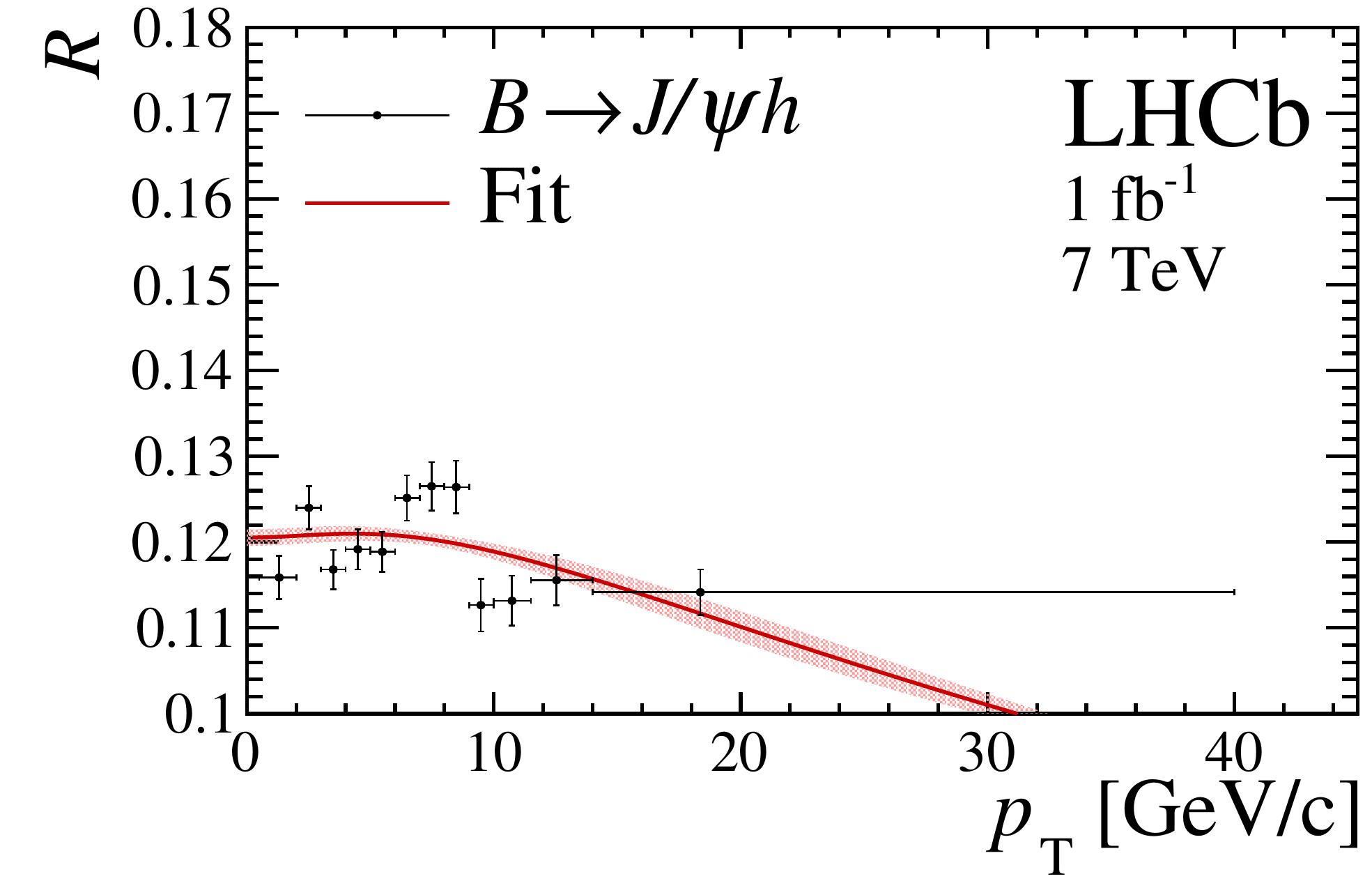}
     \includegraphics[width =0.32\textwidth ]{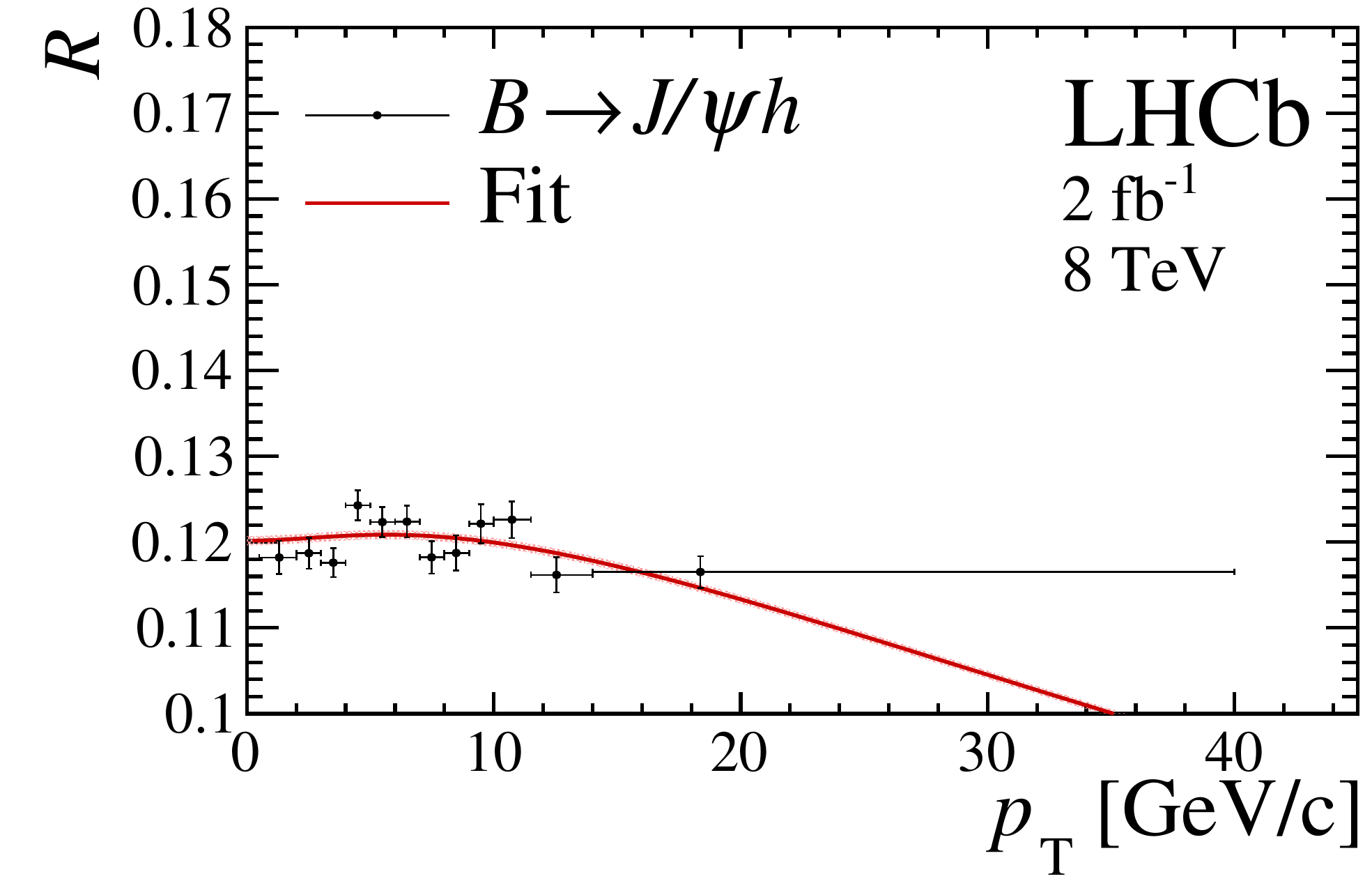}
     \includegraphics[width =0.32\textwidth ]{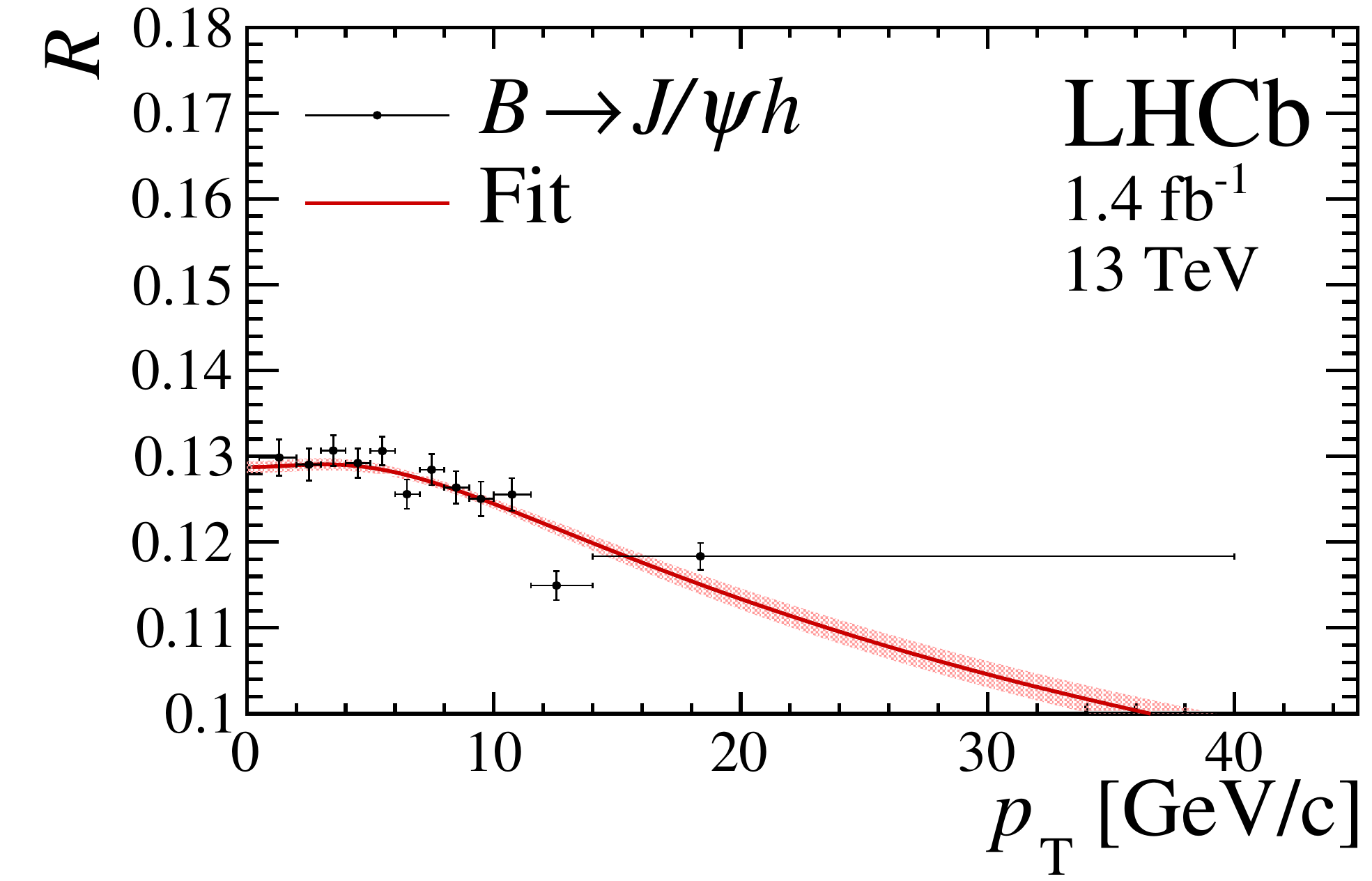}\\
     \includegraphics[width =0.32\textwidth ]{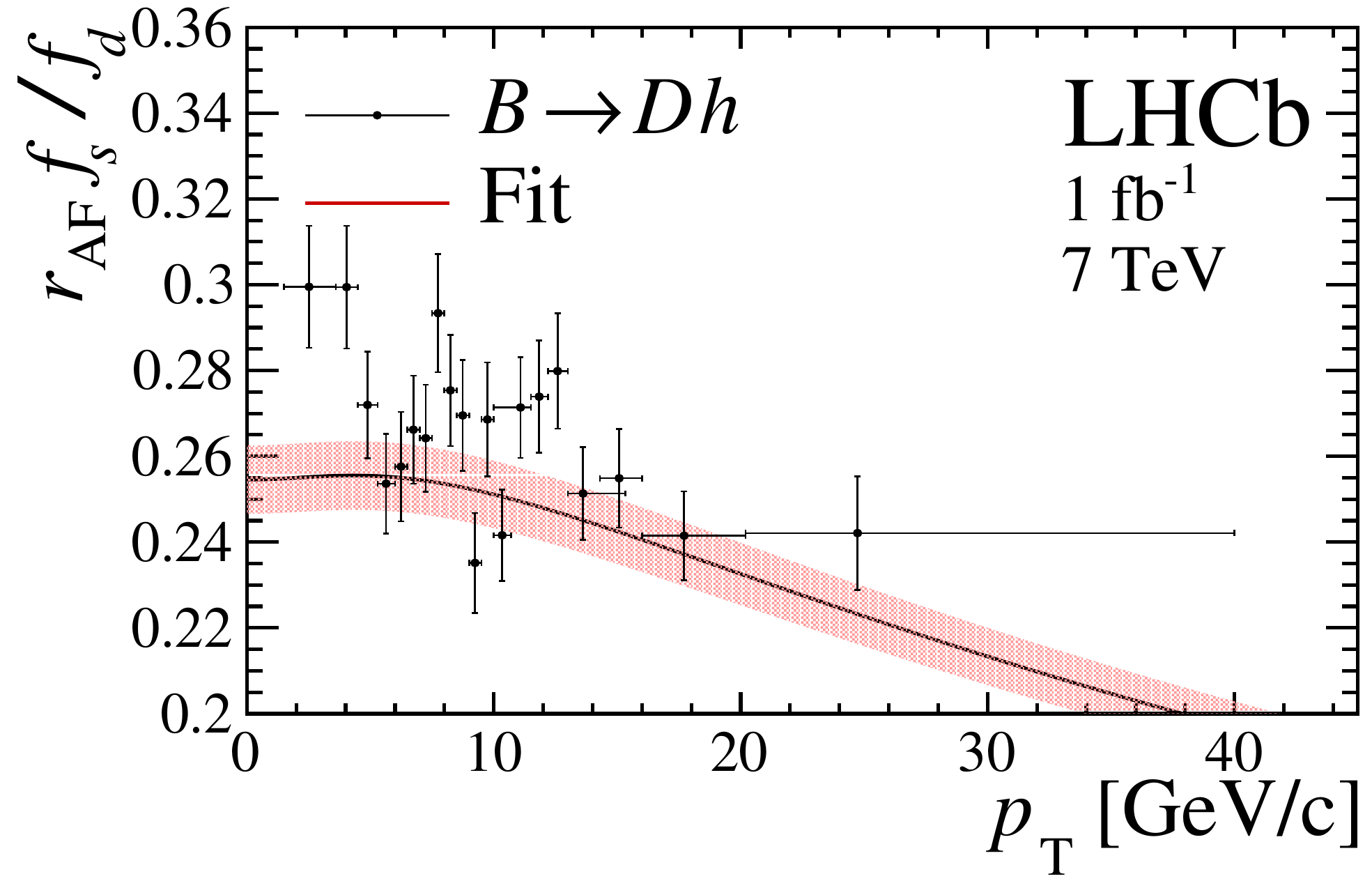}
     \includegraphics[width =0.32\textwidth ]{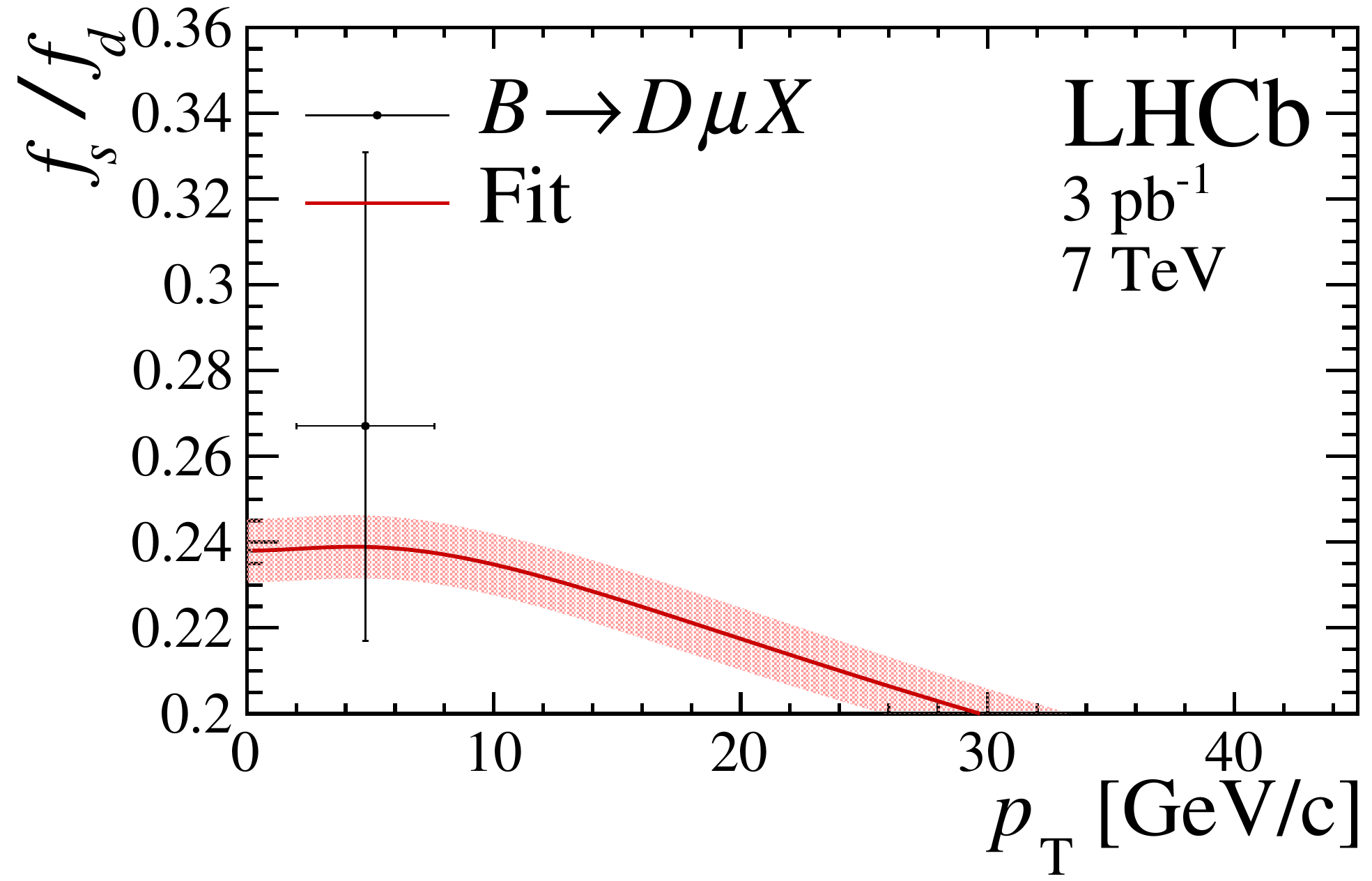}
     \includegraphics[width =0.32\textwidth ]{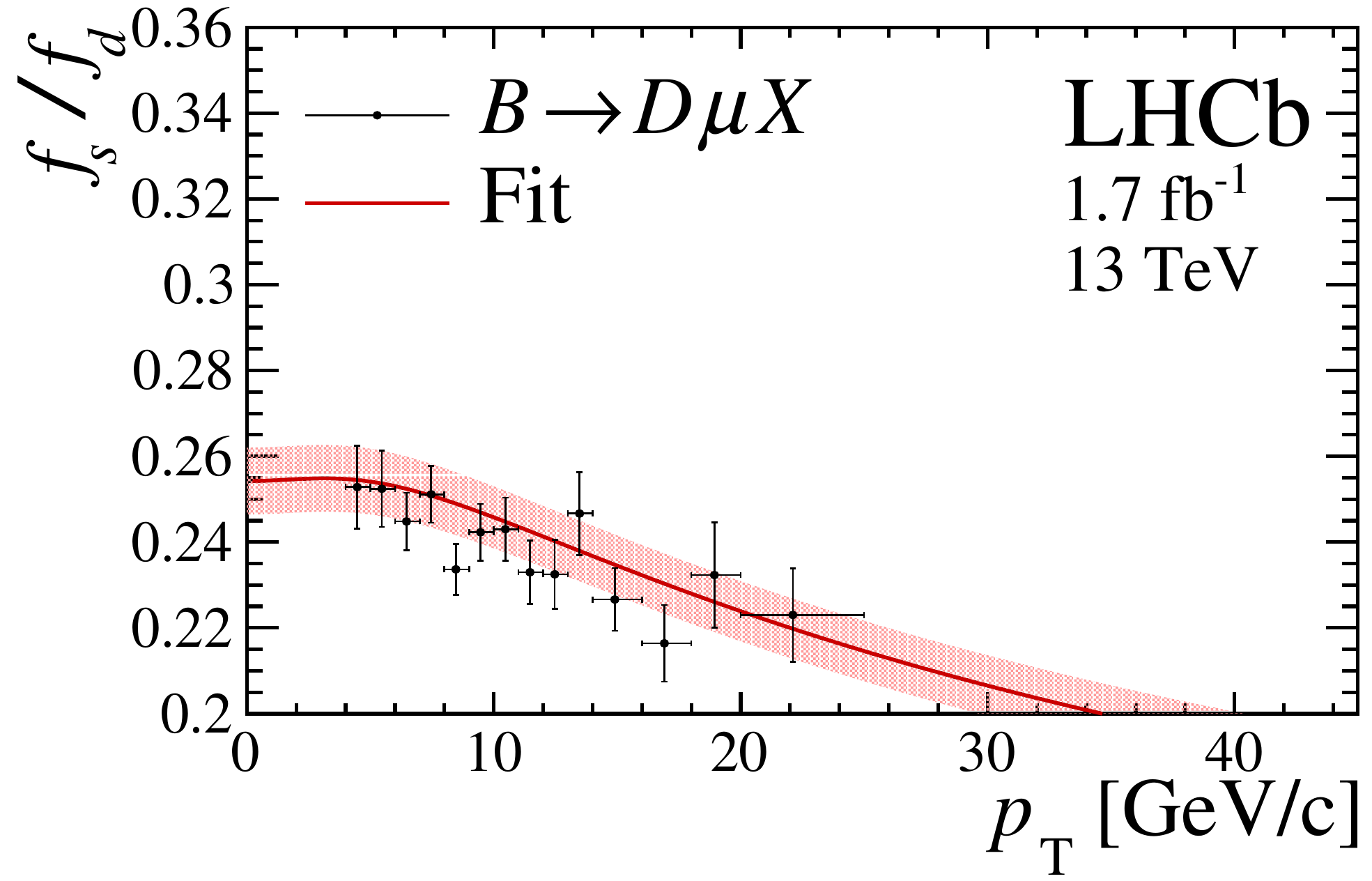}\\
     \includegraphics[width =0.32\textwidth ]{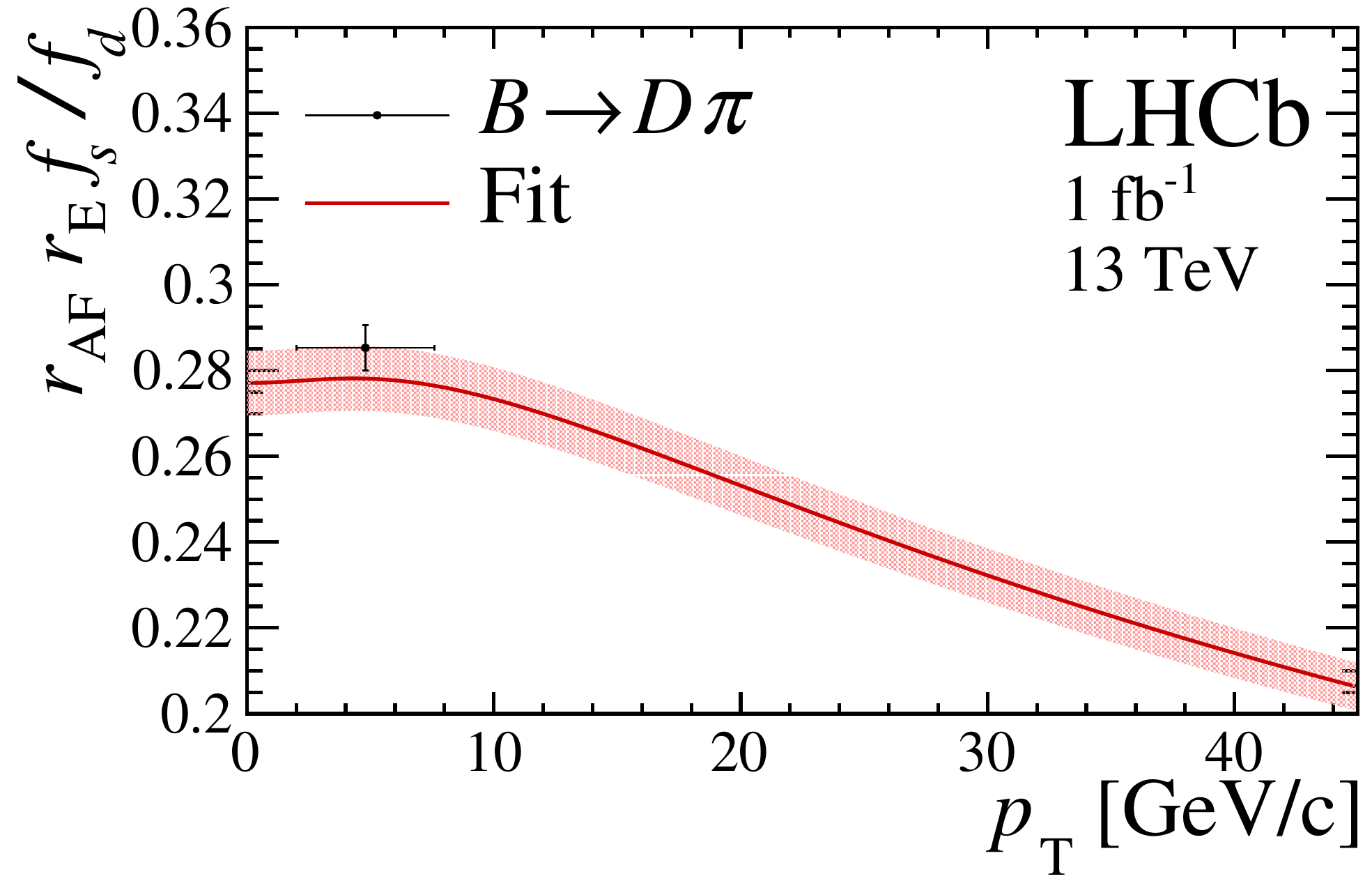}
     \includegraphics[width =0.32\textwidth ]{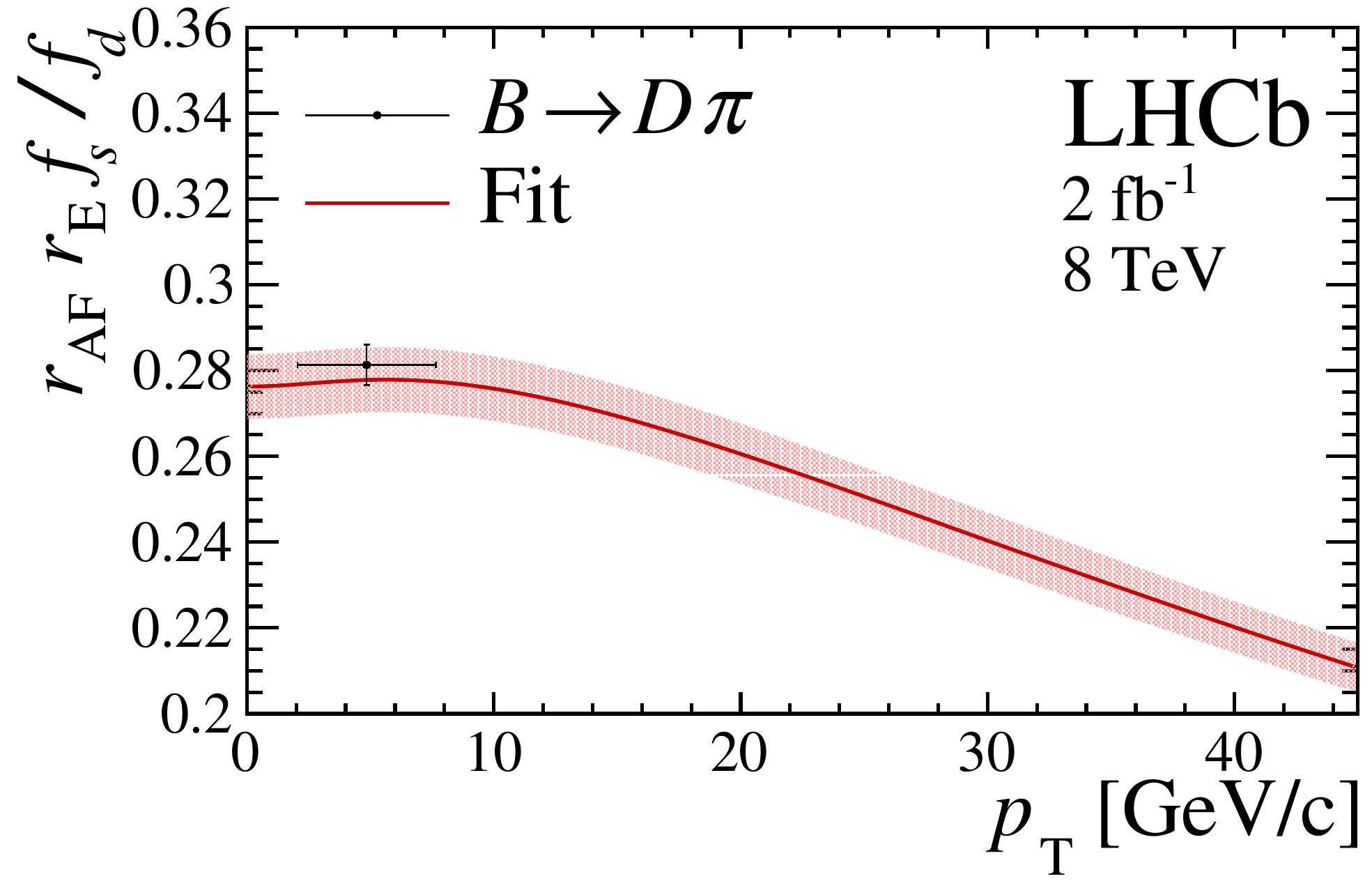}
     \includegraphics[width =0.32\textwidth ]{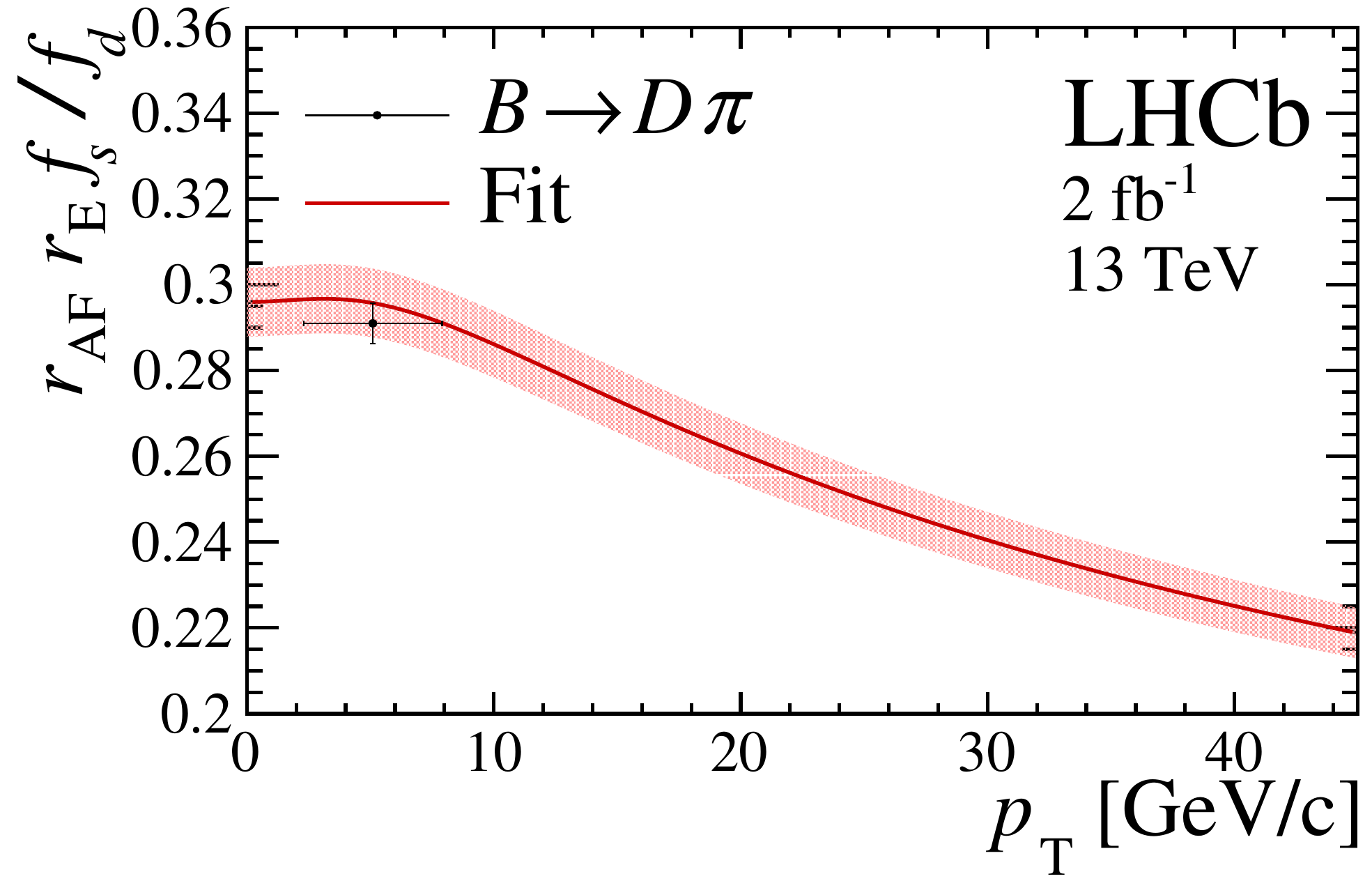}
  \caption{Measurements of \fsfd sensitive observables as a function of the \B-meson transverse momentum, 
	   \pt, overlaid with the fit function. A Tsallis-statistics inspired function is used in
	   this plot as described in the text. The scaling factors $r_{AF}$ and $r_{E}$ are defined in the text;
	   the variable \R is defined in Eq.~\ref{eq:R}. The vertical axes are zero-suppressed.
	   The uncertainties on the data points are fully independent of each other;
	   overall uncertainties for measurements in multiple \pt intervals are 
	   propagated through scaling parameters, as described in the text.
           The band associated with the fit function shows the uncertainty
	   on the post-fit function for each sample.
	   }\label{fig:tsallisfunction}
\end{figure}

The \pt distribution of produced mesons 
is often described through a function inspired by the Tsallis statistics~\cite{Tsallis:1987eu,Tsallis:1999nq}. 
Examples of this use can be found in Refs.\cite{Abelev:2006cs,Adare:2011vy,Aamodt:2011zj,Khachatryan:2011tm,Aad:2010ac}.
In particular, factoring out the pseudorapidity-dependent part, 
this function is often written as 
\begin{equation}
 \frac{d  N}{d\pt} = C  \frac{(n-1)(n-2)}{n T (nT + Mc^2 (n-2))} \pt \left[1 + \frac{  \sqrt{M^2c^4 + p^2_T c^2} - M c^2 }{n T }  \right]^{-n}\quad,
\end{equation}
where $M$ is the mass of the meson, $n$ and $T$ are parameters linked to the Tsallis statistics, and $C$ is a normalisation constant. 
An attempt has been made to describe the data with a ratio of two such Tsallis functions. 
Reasonable agreement, albeit with large fit instabilities due to parametrisation ambiguities, 
is obtained when considering the same value for the $T$ parameter 
for the \Bs and \Bd mesons, and with the $n$ differing by a factor of 0.9 between \Bs and \Bd mesons.
The results of this fit tantalisingly reproduce the stabilisation, or even decrease, of the \fsfd seen in the data at low \pt values, and are reported in Fig.~\ref{fig:tsallisfunction}.
The branching fractions obtained with this parametrisation are in agreement with the default fit, 
but have larger uncertainties due to the fit instability.

\section{Conclusion}
\label{sec:Conclusion}
In conclusion, this paper presents a precise measurement of the ratio of \Bs and \Bd fragmentation fractions, \fsfd, as a function of $\proton\proton$ centre-of-mass energy \sqs and \B-meson \pt,
from the combined analysis of LHCb measurements,
significantly reducing the uncertainty with respect to the individual measurements.
A significant dependence of \fsfd on \sqs and \pt, described by linear functions, is observed.
The integrated \fsfd values at the three energies, in the fiducial region of the measurements, are
\begin{eqnarray*}
    \fsfd\,(7\tev) &=&  \fsfdintseventev ~, \nonumber \\
    \fsfd\,(8\tev) &=&  \fsfdinteighttev ~, \\
    \fsfd\,(13\tev) &=& \fsfdintthirteentev ~, \nonumber
\end{eqnarray*}
and the ratio of the 13 to $8\tev$ results is
\begin{eqnarray*}
\frac{\fsfd\,(13\tev)}{\fsfd\,(8\tev)} &=&  \text{$1.065 \pm 0.007$} \quad. 
\end{eqnarray*}

Precise measurements of the \bsjpsiphi and \BsDspi branching fractions, 
 \begin{align*}
      \BF(\bsjpsiphi) & = \BRBsjpsiphi   \quad, \\
 \BF(\BsDspi) & = \BRBsDspivalue\quad,
 \end{align*}
are also obtained, 
halving their uncertainties with respect to previous world averages.
Finally, previous LHCb measurements of \Bs branching fractions are updated,
strongly reducing their normalisation-related uncertainties 
and better constraining possible contributions from physics beyond the SM.

\clearpage

\section*{Acknowledgements}
\noindent We express our gratitude to our colleagues in the CERN
accelerator departments for the excellent performance of the LHC. We
thank the technical and administrative staff at the LHCb
institutes.
We acknowledge support from CERN and from the national agencies:
CAPES, CNPq, FAPERJ and FINEP (Brazil); 
MOST and NSFC (China); 
CNRS/IN2P3 (France); 
BMBF, DFG and MPG (Germany); 
INFN (Italy); 
NWO (Netherlands); 
MNiSW and NCN (Poland); 
MEN/IFA (Romania); 
MSHE (Russia); 
MICINN (Spain); 
SNSF and SER (Switzerland); 
NASU (Ukraine); 
STFC (United Kingdom); 
DOE NP and NSF (USA).
We acknowledge the computing resources that are provided by CERN, IN2P3
(France), KIT and DESY (Germany), INFN (Italy), SURF (Netherlands),
PIC (Spain), GridPP (United Kingdom), RRCKI and Yandex
LLC (Russia), CSCS (Switzerland), IFIN-HH (Romania), CBPF (Brazil),
PL-GRID (Poland) and NERSC (USA).
We are indebted to the communities behind the multiple open-source
software packages on which we depend.
Individual groups or members have received support from
ARC and ARDC (Australia);
AvH Foundation (Germany);
EPLANET, Marie Sk\l{}odowska-Curie Actions and ERC (European Union);
A*MIDEX, ANR, Labex P2IO and OCEVU, and R\'{e}gion Auvergne-Rh\^{o}ne-Alpes (France);
Key Research Program of Frontier Sciences of CAS, CAS PIFI, CAS CCEPP, 
Fundamental Research Funds for the Central Universities, 
and Sci. \& Tech. Program of Guangzhou (China);
RFBR, RSF and Yandex LLC (Russia);
GVA, XuntaGal and GENCAT (Spain);
the Leverhulme Trust, the Royal Society
 and UKRI (United Kingdom).

\clearpage

\appendix
\section{Supplemental material}
\label{sec:supplemental}

Full results on the parameters for the default fit and the fit without external theory constraints are presented in Tables~\ref{tab:fitresults}
and~\ref{tab:fitresults_nohad}, respectively. 
The correlation matrices of the results for these fits also shown in Tables~\ref{tab:correlation} and~\ref{tab:correlation_nohad}, respectively.
This is important when considering the results at different energies, since they are correlated among each other.

The listed parameters are: 
\begin{itemize}
 \item[-] $a$ and $b$ are the intercept and slope of the transverse-momentum-dependent functions 
at the three center-of-mass energies;
 \item[-] $r_{\rm{AF}}$ and $r_{\rm{E}}$ are the scaling parameters with respect to the theoretical inputs;
 \item[-] $S_1$ is the parameter propagating the correlated systematic uncertainty due to external parameters;
 \item[-] $S_2$, $S_3$, and $S_4$ are the parameters propagating experimental systematic uncertainties.
 \item[-] $F_R$ is the ratio of the \bsjpsiphi to \bujpsik branching fractions, as detailed in the text.
\end{itemize}

\def\interceptseventev{$a(7\tev)$}
\def\slopeseventev{$b(7\tev)$}
\def\SystCorr{ $S_1$} 
\def\SystHadExpRone{$S_2$}
\def\SystHadNANF{$r_{\rm{AF}}$}
\def\AR{\FR}
\def\intercepteighttev{$a(8\tev)$}
\def\slopeeighttev{$b(8\tev)$}
\def\interceptonethreetev{$a(13\tev)$} 
\def\slopeonethreetev{$b(13\tev)$}
\def\SystSemiCorr{$S_3$} 
\def\SystSemitwoUncorr{$S_4$}
\def\SystHadNE{$r_{\rm{E}}$}

\begin{table}[bp]
\caption{Output parameters of the default fit to the data.}\label{tab:fitresults}
\centering
\begin{tabular}{rrrrrrrrrrrrrrrrrr}
\toprule
 \interceptseventev  	&  $0.244 \pm 0.008$ \\ 
\slopeseventev  	&  $(-10.3 \pm 2.7)\times 10^{-4}$ \\ 
\SystCorr  	&  $1.009 \pm 0.026$ \\ 
\SystHadExpRone  	&  $1.030 \pm 0.028$ \\ 
\SystHadNANF  	&  $1.082 \pm 0.032$ \\ 
\AR  	&  $0.505 \pm 0.016$ \\ 
\intercepteighttev  	&  $0.240 \pm 0.008$ \\ 
\slopeeighttev  	&  $(-3.5 \pm 2.3)\times 10^{-4}$ \\ 
\interceptonethreetev  	&  $0.263 \pm 0.008$ \\ 
\slopeonethreetev  	&  $(-17.6 \pm 2.1)\times 10^{-4}$ \\ 
\SystSemiCorr  	&  $0.997 \pm 0.008$ \\ 
\SystSemitwoUncorr  	&  $0.977 \pm 0.021$ \\ 
\SystHadNE  	&  $1.071 \pm 0.030$ \\ 
  \bottomrule
\end{tabular}
\end{table}

\begin{table}[tbp]
\caption{Output parameters of the fit to the data without external theory constraints.}\label{tab:fitresults_nohad}
\centering
\begin{tabular}{rrrrrrrrrrrrrrrrrr}
\toprule
 \interceptseventev  	&  $0.238 \pm 0.008$ \\ 
\slopeseventev  	&  $(-10.3 \pm 2.7)\times 10^{-4}$ \\ 
\SystCorr  	&  $1.000 \pm 0.026$ \\ 
\SystHadExpRone  	&  $1.00 \pm 0.04$ \\ 
\SystHadNANF  	&  $1.16 \pm 0.06$ \\ 
\AR  	&  $0.517 \pm 0.017$ \\ 
\intercepteighttev  	&  $0.234 \pm 0.008$ \\ 
\slopeeighttev  	&  $(-3.3 \pm 2.3)\times 10^{-4}$ \\ 
\interceptonethreetev  	&  $0.256 \pm 0.009$ \\ 
\slopeonethreetev  	&  $(-16.9 \pm 2.0)\times 10^{-4}$ \\ 
\SystSemiCorr  	&  $1.000 \pm 0.009$ \\ 
\SystSemitwoUncorr  	&  $0.998 \pm 0.023$ \\ 
\SystHadNE  	&  $1.04 \pm 0.04$ \\ 
  \bottomrule
\end{tabular}
\end{table}

\begin{table}[tbp]
\caption{Output correlation matrix of the default fit versus \pt.}\label{tab:correlation}
\scalebox{0.6}{
\begin{tabular}{rrrrrrrrrrrrrrrrrr}
\toprule
 	 & \interceptseventev	 & \slopeseventev	 & \SystCorr	 & \SystHadExpRone	 & \SystHadNANF	 & \AR	 & \intercepteighttev	 & \slopeeighttev	 & \interceptonethreetev	 & \slopeonethreetev	 & \SystSemiCorr	 & \SystSemitwoUncorr	 & \SystHadNE\\ \midrule 
\interceptseventev 	 & $1.000$ 	 & $-0.360$ 	 & $-0.589$ 	 & $-0.185$ 	 & $-0.318$ 	 & $-0.955$ 	 & $0.925$ 	 & $-0.046$ 	 & $0.933$ 	 & $-0.314$ 	 & $-0.223$ 	 & $-0.645$ 	 & $-0.198$ \\ 
\slopeseventev 	 &  	 & $1.000$ 	 & $0.067$ 	 & $-0.045$ 	 & $-0.003$ 	 & $0.131$ 	 & $-0.129$ 	 & $0.010$ 	 & $-0.130$ 	 & $0.048$ 	 & $0.034$ 	 & $0.097$ 	 & $0.109$ \\ 
\SystCorr 	 &  	 &  	 & $1.000$ 	 & $-0.075$ 	 & $-0.128$ 	 & $0.615$ 	 & $-0.596$ 	 & $0.029$ 	 & $-0.601$ 	 & $0.170$ 	 & $0.022$ 	 & $0.064$ 	 & $-0.079$ \\ 
\SystHadExpRone 	 &  	 &  	 &  	 & $1.000$ 	 & $-0.542$ 	 & $0.193$ 	 & $-0.184$ 	 & $0.004$ 	 & $-0.186$ 	 & $0.068$ 	 & $0.083$ 	 & $0.239$ 	 & $0.841$ \\ 
\SystHadNANF 	 &  	 &  	 &  	 &  	 & $1.000$ 	 & $0.328$ 	 & $-0.320$ 	 & $0.019$ 	 & $-0.322$ 	 & $0.129$ 	 & $0.142$ 	 & $0.410$ 	 & $-0.569$ \\ 
\AR 	 &  	 &  	 &  	 &  	 &  	 & $1.000$ 	 & $-0.967$ 	 & $0.044$ 	 & $-0.976$ 	 & $0.326$ 	 & $0.233$ 	 & $0.676$ 	 & $0.198$ \\ 
\intercepteighttev 	 &  	 &  	 &  	 &  	 &  	 &  	 & $1.000$ 	 & $-0.257$ 	 & $0.945$ 	 & $-0.318$ 	 & $-0.226$ 	 & $-0.654$ 	 & $-0.202$ \\ 
\slopeeighttev 	 &  	 &  	 &  	 &  	 &  	 &  	 &  	 & $1.000$ 	 & $-0.046$ 	 & $0.021$ 	 & $0.010$ 	 & $0.030$ 	 & $0.030$ \\ 
\interceptonethreetev 	 &  	 &  	 &  	 &  	 &  	 &  	 &  	 &  	 & $1.000$ 	 & $-0.492$ 	 & $-0.228$ 	 & $-0.660$ 	 & $-0.202$ \\ 
\slopeonethreetev 	 &  	 &  	 &  	 &  	 &  	 &  	 &  	 &  	 &  	 & $1.000$ 	 & $0.056$ 	 & $0.161$ 	 & $0.098$ \\ 
\SystSemiCorr 	 &  	 &  	 &  	 &  	 &  	 &  	 &  	 &  	 &  	 &  	 & $1.000$ 	 & $-0.059$ 	 & $0.087$ \\ 
\SystSemitwoUncorr 	 &  	 &  	 &  	 &  	 &  	 &  	 &  	 &  	 &  	 &  	 &  	 & $1.000$ 	 & $0.251$ \\ 
\SystHadNE 	 &  	 &  	 &  	 &  	 &  	 &  	 &  	 &  	 &  	 &  	 &  	 &  	 & $1.000$ \\ 
  \end{tabular}}
\end{table}

\begin{table}[tbp]
\caption{Output correlation matrix of the fit versus \pt without theory constraints.}\label{tab:correlation_nohad}
\scalebox{0.6}{
\begin{tabular}{rrrrrrrrrrrrrrrrrr}
\toprule
 	 & \interceptseventev	 & \slopeseventev	 & \SystCorr	 & \SystHadExpRone	 & \SystHadNANF	 & \AR	 & \intercepteighttev	 & \slopeeighttev	 & \interceptonethreetev	 & \slopeonethreetev	 & \SystSemiCorr	 & \SystSemitwoUncorr	 & \SystHadNE\\ \midrule 
\interceptseventev 	 & $1.000$ 	 & $-0.343$ 	 & $-0.525$ 	 & $0.001$ 	 & $-0.367$ 	 & $-0.958$ 	 & $0.931$ 	 & $-0.049$ 	 & $0.938$ 	 & $-0.333$ 	 & $-0.257$ 	 & $-0.672$ 	 & $-0.002$ \\ 
\slopeseventev 	 &  	 & $1.000$ 	 & $0.069$ 	 & $0.000$ 	 & $-0.032$ 	 & $0.125$ 	 & $-0.123$ 	 & $0.011$ 	 & $-0.124$ 	 & $0.048$ 	 & $0.034$ 	 & $0.088$ 	 & $0.111$ \\ 
\SystCorr 	 &  	 &  	 & $1.000$ 	 & $0.000$ 	 & $-0.166$ 	 & $0.548$ 	 & $-0.531$ 	 & $0.027$ 	 & $-0.536$ 	 & $0.150$ 	 & $0.003$ 	 & $0.007$ 	 & $0.000$ \\ 
\SystHadExpRone 	 &  	 &  	 &  	 & $1.000$ 	 & $-0.768$ 	 & $-0.001$ 	 & $0.001$ 	 & $-0.000$ 	 & $0.001$ 	 & $-0.000$ 	 & $-0.000$ 	 & $-0.001$ 	 & $0.920$ \\ 
\SystHadNANF 	 &  	 &  	 &  	 &  	 & $1.000$ 	 & $0.378$ 	 & $-0.367$ 	 & $0.019$ 	 & $-0.370$ 	 & $0.152$ 	 & $0.178$ 	 & $0.467$ 	 & $-0.787$ \\ 
\AR 	 &  	 &  	 &  	 &  	 &  	 & $1.000$ 	 & $-0.970$ 	 & $0.048$ 	 & $-0.978$ 	 & $0.343$ 	 & $0.267$ 	 & $0.701$ 	 & $-0.004$ \\ 
\intercepteighttev 	 &  	 &  	 &  	 &  	 &  	 &  	 & $1.000$ 	 & $-0.252$ 	 & $0.949$ 	 & $-0.336$ 	 & $-0.260$ 	 & $-0.680$ 	 & $-0.005$ \\ 
\slopeeighttev 	 &  	 &  	 &  	 &  	 &  	 &  	 &  	 & $1.000$ 	 & $-0.049$ 	 & $0.023$ 	 & $0.013$ 	 & $0.034$ 	 & $0.018$ \\ 
\interceptonethreetev 	 &  	 &  	 &  	 &  	 &  	 &  	 &  	 &  	 & $1.000$ 	 & $-0.502$ 	 & $-0.262$ 	 & $-0.686$ 	 & $-0.004$ \\ 
\slopeonethreetev 	 &  	 &  	 &  	 &  	 &  	 &  	 &  	 &  	 &  	 & $1.000$ 	 & $0.073$ 	 & $0.191$ 	 & $0.018$ \\ 
\SystSemiCorr 	 &  	 &  	 &  	 &  	 &  	 &  	 &  	 &  	 &  	 &  	 & $1.000$ 	 & $0.004$ 	 & $-0.000$ \\ 
\SystSemitwoUncorr 	 &  	 &  	 &  	 &  	 &  	 &  	 &  	 &  	 &  	 &  	 &  	 & $1.000$ 	 & $-0.001$ \\ 
\SystHadNE 	 &  	 &  	 &  	 &  	 &  	 &  	 &  	 &  	 &  	 &  	 &  	 &  	 & $1.000$ \\ 
  \end{tabular}}
\end{table}

\clearpage

\addcontentsline{toc}{section}{References}
\bibliographystyle{LHCb}
\bibliography{main,standard,LHCb-PAPER,LHCb-CONF,LHCb-DP,LHCb-TDR}

\newpage
\centerline
{\large\bf LHCb collaboration}
\begin
{flushleft}
\small
R.~Aaij$^{32}$,
C.~Abell{\'a}n~Beteta$^{50}$,
T.~Ackernley$^{60}$,
B.~Adeva$^{46}$,
M.~Adinolfi$^{54}$,
H.~Afsharnia$^{9}$,
C.A.~Aidala$^{85}$,
S.~Aiola$^{25}$,
Z.~Ajaltouni$^{9}$,
S.~Akar$^{65}$
% ...

{\footnotesize \it

% ...
$^{9}$Universit{\'e} Clermont Auvergne, CNRS/IN2P3, LPC, Clermont-Ferrand, France\\
% ...
$^{25}$INFN Sezione di Milano, Milano, Italy\\
% ...
$^{32}$Nikhef National Institute for Subatomic Physics, Amsterdam, Netherlands\\
% ...
$^{46}$Instituto Galego de F{\'\i}sica de Altas Enerx{\'\i}as (IGFAE), Universidade de Santiago de Compostela, Santiago de Compostela, Spain\\
% ...
$^{50}$Physik-Institut, Universit{\"a}t Z{\"u}rich, Z{\"u}rich, Switzerland\\
% ...
$^{54}$H.H. Wills Physics Laboratory, University of Bristol, Bristol, United Kingdom\\
% ...
$^{60}$Oliver Lodge Laboratory, University of Liverpool, Liverpool, United Kingdom\\
% ...
$^{65}$University of Cincinnati, Cincinnati, OH, United States\\
% ...
$^{68}$Syracuse University, Syracuse, NY, United States\\
% ...
$^{85}$University of Michigan, Ann Arbor, United States, associated to $^{68}$\\
\bigskip
% ...
\medskip
}
\end{flushleft} 

\end{document}